\documentclass[preprint2]{aastex}

\usepackage{amsmath}
\usepackage{epstopdf}
\usepackage[latin1]{inputenc}
\slugcomment{Submitted to Astrophysical Journal}

\shorttitle{The stickiness of micrometer-sized water-ice particles}
\shortauthors{B. Gundlach and J. Blum}

\begin{document}

\title{The stickiness of micrometer-sized water-ice particles}

\author{B. Gundlach and J. Blum}
\affil{Institut für Geophysik und extraterrestrische Physik, Technische Universität Braunschweig,
Mendelssohnstr. 3, D-38106 Braunschweig, Germany}

\begin{abstract}
Water ice is one of the most abundant materials in dense molecular clouds and in the outer reaches of protoplanetary disks. In contrast to other materials (e.g., silicates) water ice is assumed to be stickier due to its higher specific surface energy, leading to faster or more efficient growth in mutual collisions. However, experiments investigating the stickiness of water ice have been scarce, particularly in the astrophysically relevant micrometer-size region and at low temperatures. In this work, we present an experimental setup to grow aggregates composed of $\mathrm{\mu}$m-sized water-ice particles, which we used to measure the sticking and erosion thresholds of the ice particles at different temperatures between $114 \, \mathrm{K}$ and $260 \, \mathrm{K}$. We show with our experiments that for low temperatures (below $\sim 210 \, \mathrm{K}$), $\mathrm{\mu}$m-sized water-ice particles stick below a threshold velocity of $9.6 \, \mathrm{m \, s^{-1}}$, which is approximately ten times higher than the sticking threshold of $\mathrm{\mu}$m-sized silica particles. Furthermore, erosion of the grown ice aggregates is observed for velocities above $15.3 \, \mathrm{m \, s^{-1}}$. A comparison of the experimentally derived sticking threshold with model predictions is performed to determine important material properties of water ice, i.e., the specific surface energy and the viscous relaxation time. Our experimental results indicate that the presence of water ice in the outer reaches of protoplanetary disks can enhance the growth of planetesimals by direct sticking of particles.
\end{abstract}


\section{Introduction}
Water ice is a fascinating material, which is of utmost importance for the physical and chemical evolution of dense molecular-cloud cores and the outer reaches of protoplanetary disks, because it is one of the most abundant species in these objects \citep[see review by][]{Dishoek2010}. Its occurrence as solid material as part of the interstellar matter is typically in (sub-)$\mathrm{\mu}$m-sized grains, which are often termed as "dust", together with other, less volatile, solids.
\par
Dust particles in general are ubiquitous in space. Wherever dust grains are present, they give rise to exciting physical processes. Their interaction with electromagnetic radiation (absorption, emission, scattering) determines the appearance and the spectral energy distribution of many astronomical objects. When subjected to high-energy particle fluxes, sputtering can reduce the sizes of dust grains, enhance the abundances of gaseous species, and deliver ions into the gas phase. On the other hand, dust grains act as condensation nuclei in regions with high densities and low temperatures. Mutual dust collisions can lead to sticking (coagulation) for low impact velocities and to shattering (fragmentation) if the collision speeds are high enough. For intermediate velocities, cratering or erosion can occur \citep{BlumWurm2008}. Depending on the temperature and the harshness of the environment, dust particles consist of highly refractory materials (e.g., oxides, metals, silicates) or exhibit a more volatile composition (e.g., organic material, ices). Interstellar dust particles in cold regions possibly possess a core-mantle-type morphology, with a refractory silicate core, an inner organic mantle, and an outer ice mantle \citep{Li1997}
\par
Dust coagulation plays an important role in the densest interstellar regions, i.e., in molecular clouds \citep{Ossenkopf1993, Weidenschilling1994, Ormel2009} and protoplanetary disks \citep{BlumWurm2008, Zsom2010, Testi2014}. Collision velocities in these regions are sufficiently low to allow two dust grains (or two small dust aggregates) to stick to one another after a collision. Evidence for dust growth in molecular clouds \citep[e.g.,][]{Steinacker2010, Pagani2010} and in protoplanetary disks \citep{Wilner2005, Rodmann2006, Testi2014} is available. In both environments, the dominating dust materials are silicates, carbonaceous materials, and ices. In contrast to silicates, for which empirical evidence for the low-velocity collision behavior exists \citep{BlumWurm2008,Guettler2010}, the collision properties of carbonaceous materials and ices (and, in particular, of the most abundant water ice) are mainly unknown. Coagulation models \citep{Wada2007, Wada2008, Wada2009, Wettlaufer2010} assume a higher upper energy limit for sticking of ice particles than for silicates \citep[for the latter, see][]{Poppe2000}, but empirical evidence for this conjecture is scarce. The only supporting fact for an enhanced stickiness of water ice with respect to silicates is the at least tenfold higher surface energy of water ice \citep[$0.19 \, \mathrm{J \, m^{-2}}$;][]{Gundlach2011b} in comparison to silicates \citep[$0.02 \, \mathrm{J \, m^{-2}}$;][]{BlumWurm2000, Gundlach2011b}. In addition, the work by \citet{Aumatelletal2014} suggests an even higher specific surface energy of water ice of $0.37 \, \mathrm{J \, m^{-2}}$.
\par
Dust collisions can also lead to fragmentation of dust aggregates in molecular clouds \citep{Ormel2009} if the collision velocities exceed the fragmentation threshold for aggregates, which is expected for velocities in the range of $\sim 1 \, - \, 100 \, \mathrm{m \, s^{-1}}$, depending on grain size and material. As in the case of coagulation, this effect has been investigated in the laboratory for silicates \citep{BlumWurm2008, Guettler2010, Blumetal2014} but not for ices. At very high collision velocities, a variety of impact experiments of (small) projectiles into (large) targets exist, even for ice at cryogenic temperatures \citep[see, e.g.,][and references therein]{Koschny2001}. At these impact velocities, crater formation with an excessive outflow of ejecta is generally observed.
\par
As outlined above, we are still lacking quantitative data on the coagulation or fragmentation behavior of microscopic water-ice particles and aggregates thereof under the conditions found in molecular cloud cores and protoplanetary disks. For macroscopic ice particles (centimeter to decimeter sizes), laboratory experiments have been performed by \citet{Hatzes1988}, \citet{Hatzes1991}, \citet{Supulver1997}, \citet{Heisselmann2010}, and Hill et al. (2014, sumitted to A$\&$A). Their findings agree inasmuch that particles consisting of water ice do not stick to one another, even at velocities below $0.1 \, \mathrm{mm \, s^{-1}}$. However, a frost layer increases the sticking threshold by some sort of "Velcro" effect. It is, however, questionable whether these experiments can be used to derive the collision properties of microscopically small water-ice particles.
\par
In this work, we present a novel experimental setup, which is used to grow ice aggregates composed of $\mathrm{\mu}$m-sized water-ice particles at low temperatures (see Sect. \ref{Experimental setup and procedure}). With this experiment, we are able to determine the sticking and the erosion thresholds of $\mathrm{\mu}$m-sized water-ice particles at different temperatures. In Sect \ref{the_ice_particles}, we characterize our $\mathrm{\mu}$m-sized water-ice particles by discussing their size distribution, the specific surface energy of water ice and the temperatures of the particles before impact. The obtained experimental results are then presented in Sect. \ref{Experimental results} and compared with a collision model in Sect. \ref{Comparion with model}. This comparison is used to deduce the specific surface energy and the viscous relaxation time of the $\mathrm{\mu}$m-sized water-ice particles. In Sect. \ref{Discussion and Outlook}, the obtained results are discussed and a short outlook on future experiments is given. Finally, Sect. \ref{Conclusions} concludes this manuscript by reviewing the main results of this work.

\section{Experimental setup and procedure}\label{Experimental setup and procedure}

\subsection{Experimental setup}\label{Experimental setup}

\begin{figure*}[t!]
\centering
\includegraphics[angle=0,width=1\textwidth]{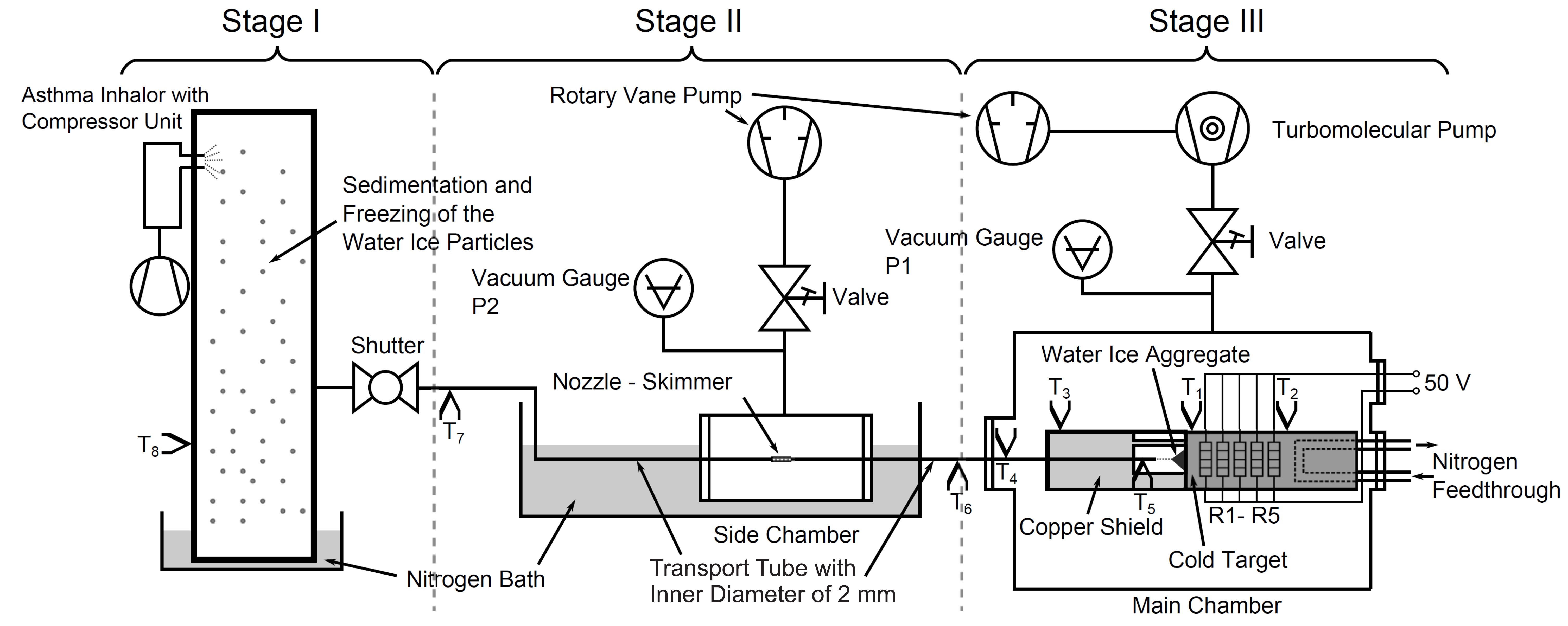}
\caption{Sketch of the experimental setup. The experiment is divided into three different stages (I-III) to ease the discussion of the ice-particle temperature during the experimental passage (see Sect. \ref{Temperature of the ice particles before collision}).}
\label{experimental_setup_skectch}
\end{figure*}

\begin{figure*}[t!]
\centering
\includegraphics[angle=0,width=0.75\textwidth]{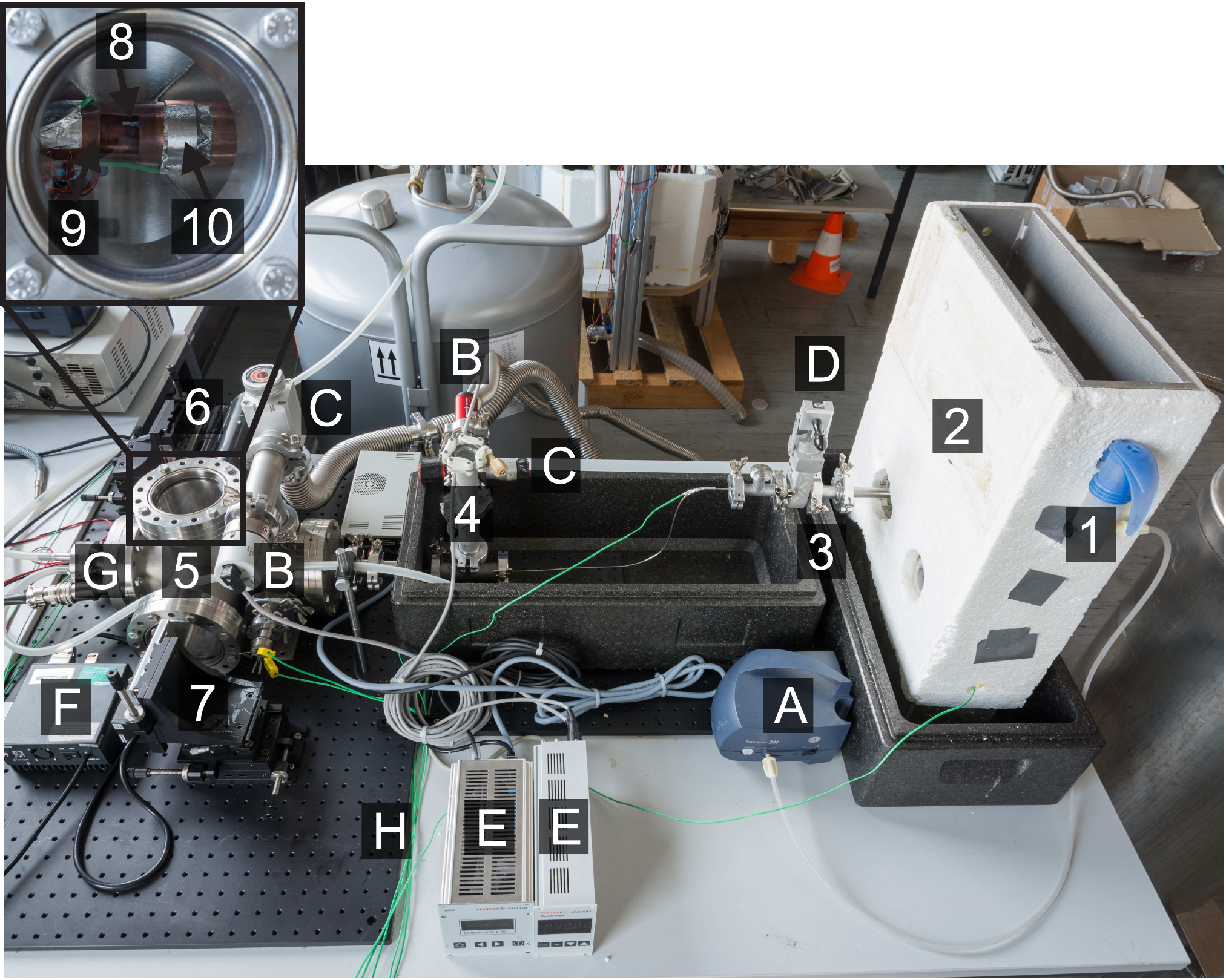}
\caption{Photograph of the experimental setup to determine the collision properties of $\mathrm{\mu}$m-sized water-ice particles. The water-ice particles are produced by spraying distilled water (1) into a cold sedimentation chamber (2). During sedimentation, the droplets slowly freeze to water-ice particles, which are then extracted (3) by a two-stage vacuum system (4 and 5). In the second vacuum chamber, an ice jet forms, which leads to collisions between the $\mathrm{\mu}$m-sized water-ice particles and a cryogenically cooled target. The collisions are observed by a long-distance microscope (6), which detects the particle-induced forward-scattered light of a pulsed laser beam (7). The inset shows the nozzle (8) used to produce the ice-particle jet and the cryogenically cooled target (9) seen from top of the second vacuum chamber through a glass window. The nozzle is surrounded by the cooling system (10) to ensure cryogenic temperatures. Additional equipment shown in the picture are the compressor for the inhalator (A), pressure sensors (B), valves (C), a shutter (D), pressure-sensor displays (E), a laser control unit (F), the target cooling system (G), and wiring for the temperature sensors (H), respectively.}
\label{experimental_setup}
\end{figure*}

We have constructed an experimental setup (see Fig. \ref{experimental_setup_skectch} for a sketch and Fig. \ref{experimental_setup} for a photograph of the experimental setup) to study the collision properties of $\mathrm{\mu}$m-sized water-ice particles in the velocity range between $\sim 1 \, \mathrm{m \, s^{-1}}$ and $\sim 150 \, \mathrm{m \, s^{-1}}$. The $\mathrm{\mu}$m-sized water-ice particles are produced by spraying distilled water with an inhalator (labeled 1 in Fig. \ref{experimental_setup}) into a sedimentation chamber (2). The sedimentation chamber is actively cooled by liquid nitrogen to temperatures between $110 \, \mathrm{K}$ and $198 \, \mathrm{K}$. Thus, freezing of the water droplets occurs during sedimentation inside the cold sedimentation chamber (see Sect. \ref{Temperature of the ice particles before collision}), leading to crystalline, spherical water-ice particles. The produced particles are then extracted (3) by the suction provided by two connected vacuum chambers (4 and 5). The first vacuum chamber (4; typical pressure: $50 \, \mathrm{mbar}$ - $200 \, \mathrm{mbar}$) is used to separate the ice particles from the gas by a nozzle-skimmer arrangement. In the second vacuum chamber (5; typical pressure: $0.1 \, \mathrm{mbar}$ - $50 \, \mathrm{mbar}$), a collimated ice-particle jet is formed. The ice-particle jet is directed onto a cryogenically cooled target (temperature: $\sim 80 \, \mathrm{K}$). The collisions between the ice particles and the target are observed by a long-distance microscope (6) using an illumination by a pulsed laser beam in forward scattering (7). To prevent the laser light from entering the microscope, it is equipped with a light trap at the center of its entry lens. Thus, only the forward-scattered light of the particles within a scattering-angle range between $4.3^\circ$ and $13.2^\circ$ can reach the microscope camera.
\par
\begin{figure}[t!]
\centering
\includegraphics[angle=0,width=1\columnwidth]{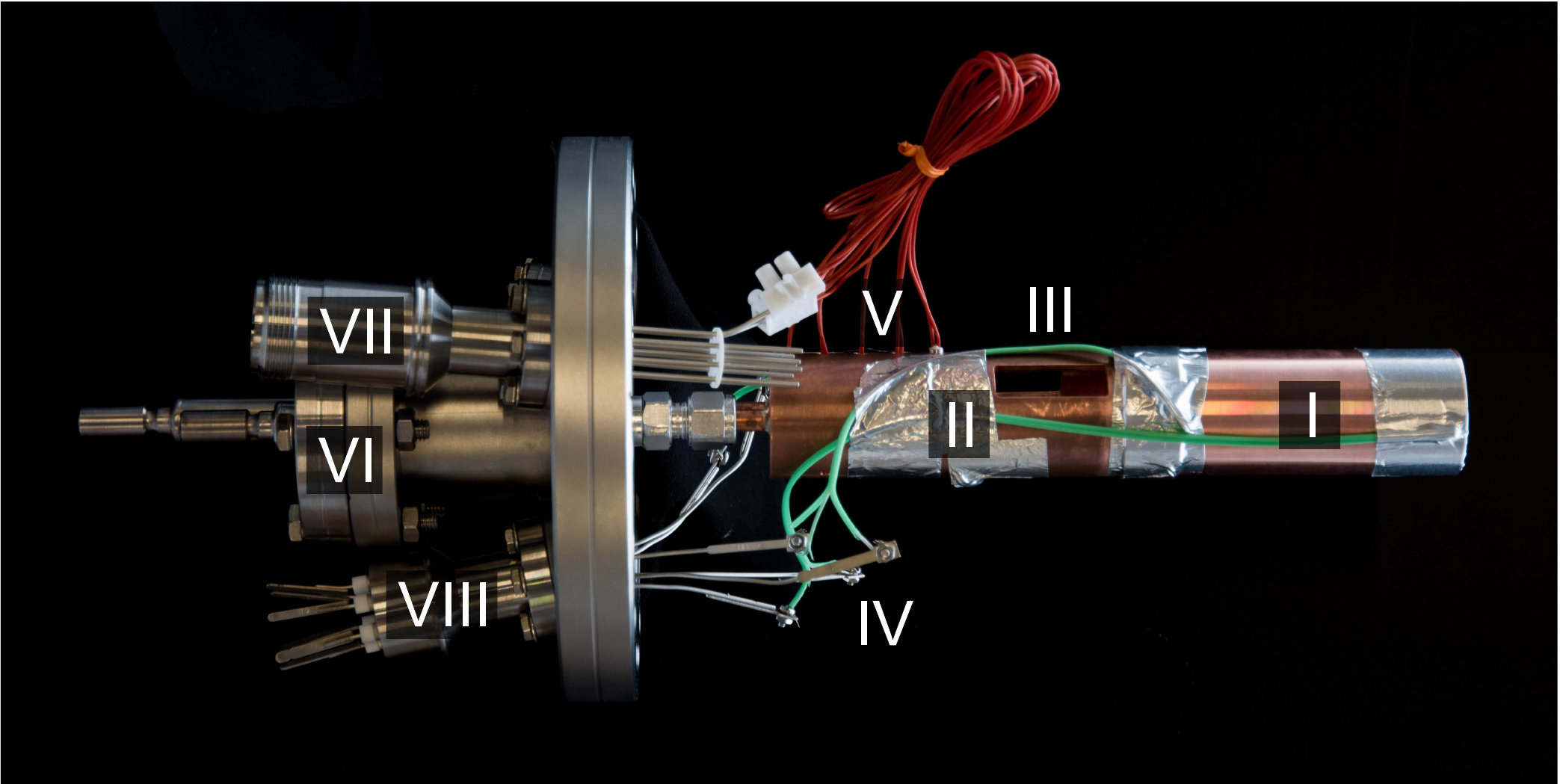}
\caption{Image of the cooling system before implementation into the experimental setup. Using liquid nitrogen, the cooling system can be actively cooled to temperatures between 100 K and 273 K. The heating cartridges allow to warm up the cooling system from cryogenic temperatures to room temperature within approximately ten minutes. The parts shown are a copper cooling shroud (I), the cold target (II), a window for observation (III), temperature sensors (IV), heating cartridges (V), the liquid-nitrogen feedthrough (VI), the electrical-power feedthrough (VII), and the temperature-sensor feedthrough (VIII), respectively.}
\label{cooling_system}
\end{figure}

The inset in Fig. \ref{experimental_setup} shows the nozzle (8) used to form the ice-particle jet and the cryogenically cooled target (9) seen from top of the second vacuum chamber through a glass window. The distance between the nozzle and the cryogenically cooled target is $2 \, \mathrm{cm}$. The nozzle is positioned inside a cooling system (10) to ensure cryogenic temperatures. The cooling system (see Fig. \ref{cooling_system}) allows active cooling to temperatures between $\sim 100 \, \mathrm{K}$ and $\sim 300 \, \mathrm{K}$ using liquid nitrogen. Heating cartridges are also included in the setup, which enable us to warm up the cooling system from cryogenic temperatures to room temperature within approximately ten minutes.
\par
The temperature sensors (H in Fig. \ref{experimental_setup} and IV in Fig. \ref{cooling_system}) are used to monitor all relevant temperatures of the experimental setup, i.e., the cooling system, the nozzle, the pipes, and the sedimentation chamber, respectively. During all experimental runs, the cooling system, the first vacuum chamber, and the sedimentation chamber were actively cooled by liquid nitrogen, which provided cryogenic temperatures. Tab. \ref{Tab_1} summarizes the measured temperatures during the experimental runs. The intervals of the measured temperatures are due to temperature fluctuations during the various experimental runs.
\par
\begin{figure}[t!]
\centering
\includegraphics[angle=0,width=1\columnwidth]{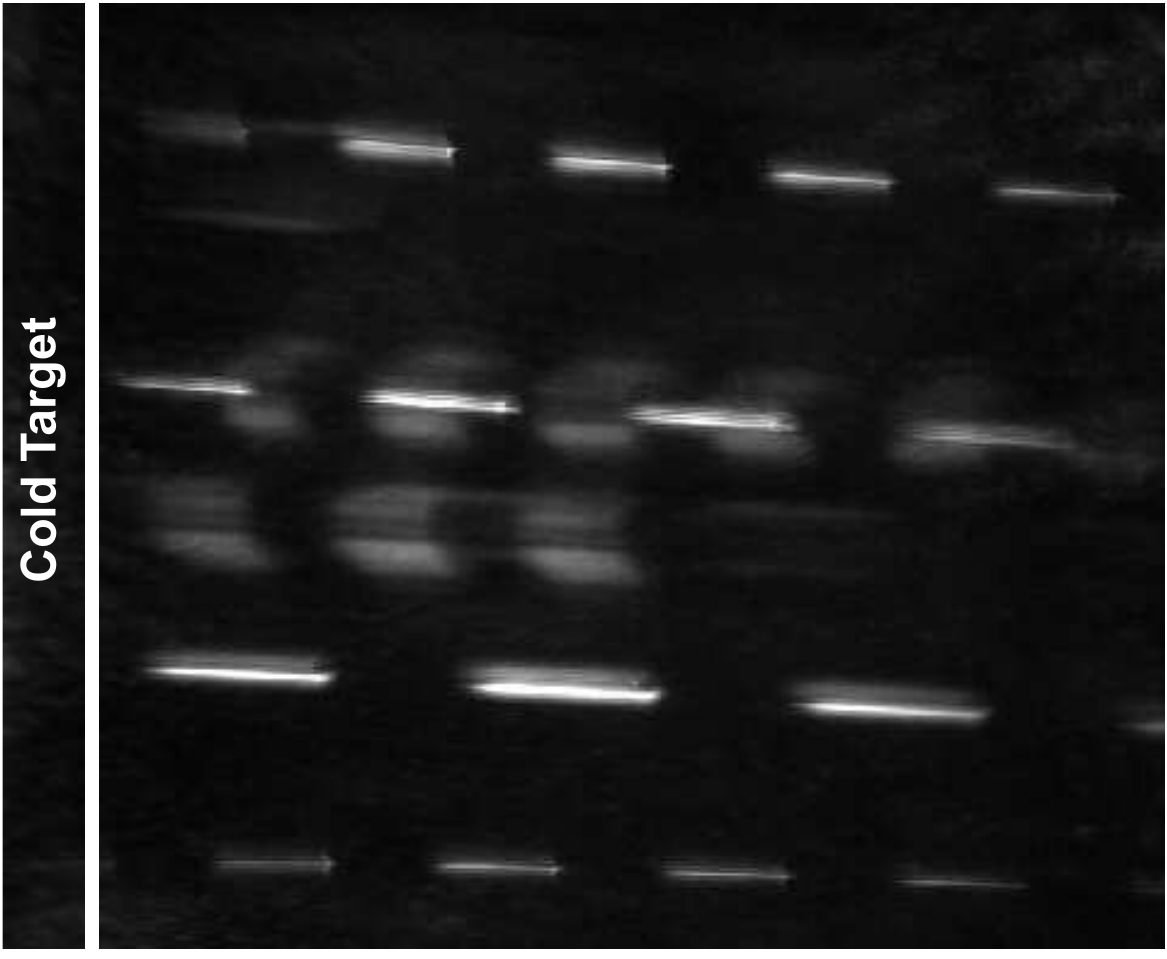}
\caption{Image containing traces of $\mathrm{\mu}$m-sized water-ice particles taken through the long-distance microscope. The ice particles are illuminated by a pulsed laser beam (frequency $59.4 \, \mathrm{kHz}$). The forward-scattered light is collected by the long-distance microscope and detected by the attached CCD camera. In this example, the velocity of the ice particles is $\sim 7 \, \mathrm{m \, s^{-1}}$. The cold target (denoted by the solid line) is not illuminated by the laser light and, thus, not visible in this image. The picture size shown is $300 \, \mathrm{\mu m}  \times 250 \, \mathrm{\mu m}$.}
\label{ice_particles}
\end{figure}

As mentioned above, the ice particles are accelerated by the pressure difference between the vacuum chambers. For typical pressures of $50 \, \mathrm{mbar}$ to $200 \, \mathrm{mbar}$ in the first vacuum chamber and $0.1 \, \mathrm{mbar}$ to $50 \, \mathrm{mbar}$ in the second vacuum chamber, the emerging particles are accelerated to velocities between $1 \, \mathrm{m \, s^{-1}}$ and $150 \, \mathrm{m \, s^{-1}}$. Fig. \ref{ice_particles} exemplarily shows $\mathrm{\mu}$m-sized water-ice particles illuminated by the pulsed laser beam and observed with the long-distance microscope (dashed lines).

\begin{deluxetable}{lccc}
\tablecolumns{8}
\tablewidth{0pc}
\tablecaption{Measured temperatures during the experimental runs. The temperatures $T_{\rm cooling\ system}$, $T_{\rm sedimentation\ chamber}$, and $T_{\rm nozzle}$ are those of the cooling system (which is identical to the temperature of the target), the sedimentation chamber, and the nozzle 2 cm away from the target, respectively.}
\tablehead{
\colhead{Experimental Run} & \colhead{$T_{\rm cooling\ system}$ [K]} & \colhead{$T_{\rm sedimentation\ chamber}$ [K]}  & \colhead{$T_{\rm nozzle}$ [K]}    }
\startdata
1 & 128 - 133  & 189 - 198 & 250 - 260 \\
2 & 123 - 140 & 150 - 154 & 235 - 243  \\
3 & 79 - 82 & 153 - 193 & 198 - 208  \\
4 & 80 - 83 & 174 - 179 & 150 - 180 \\
5 & 81 - 82 & 128 - 135 & 138 - 148 \\
6 & 80 - 84 & 110 - 116 & 114 - 120 \\
\enddata
\label{Tab_1}
\end{deluxetable}

\subsection{Experimental procedure}
In total, six experimental runs were performed at different temperatures (see Tab. \ref{Tab_1}). Before the experiments were started, the cooling system, the pipes and the sedimentation chamber were cooled by liquid nitrogen. The dispersion of the distilled water into microscopic droplets was started when the temperature of the sedimentation chamber had reached values $T_{\rm sedimentation\ chamber} < 200\,\mathrm{K}$.
\par
In the first stage of the experimental procedure (the production of the $\mathrm{\mu}$m-sized water-ice particles), the shutter between the sedimentation chamber and the vacuum chambers was closed to enhance the number density of the water-ice particles within the sedimentation chamber. In the meantime, the temperature of the cooling system and, therewith, the temperature of the nozzle (which determines the temperature of the water-ice particles when colliding with the target; see Sect. \ref{Temperature of the ice particles before collision}) was adjusted by cooling with liquid nitrogen and heating with the heating cartridges. This method enabled a temperature stability of typically $\pm 5 \, \mathrm{K}$.
\par
After the adjustment of the temperatures, the camera and the laser were switched on and the shutter was opened to extract the $\mathrm{\mu}$m-sized water-ice particles from the sedimentation chamber. During all experiments, the laser frequency was set to $59.4 \, \mathrm{kHz}$ so that equal durations of laser-on and laser-off of $8.4 \, \mathrm{\mu s}$ each were obtained. At the beginning of the experimental runs, relatively low velocities were chosen to grow ice aggregates by low-speed collisions between the $\mathrm{\mu}$m-sized water-ice particles and the cold target (see Sect. \ref{Growth and erosion of ice aggregates}).

\section{Characterization of the $\mathrm{\mu}$m-sized water-ice particles}\label{the_ice_particles}

\subsection{Size distribution}\label{Size_distr}
The size distribution of the water-ice particles was analyzed by us in previous works by using a light microscope, a long-distance microscope \citep{Gundlach2011b}, and a cryogenic-cooled scanning electron microscope \citep{Jost2013}. The three size distributions measured with the different microscopes yield mean radii of $(1.49 \pm 0.79) \, \mathrm{\mu m}$, $(1.45 \pm 0.65) \, \mathrm{\mu m}$, and $(1.47^{+0.96}_{-0.58}) \, \mathrm{\mu m}$, respectively, with the errors describing the standard deviations. The minimum and maximum observed particle radii were $0.2 \, \mathrm{\mu m}$ and $6.0 \, \mathrm{\mu m}$, respectively. The investigation of the particle size with the cryogenic-cooled scanning electron microscope has additionally shown that the $\mathrm{\mu}$m-sized water-ice particles are visibly spherical \citep[see Fig. 4 in][]{Jost2013}.

\subsection{Specific surface energy of water ice}\label{Specific surface energy}
In a previous work \citep{Gundlach2011b}, we estimated the specific surface energy of the $\mathrm{\mu}$m-sized water-ice particles by observing ice aggregates composed of $\mathrm{\mu}$m-sized water-ice particles with a long-distance microscope. When slow gravitational restructuring of the ice aggregates was observed, the critical rolling force of the water-ice particles could be measured and, therewith, the specific surface energy was derived. These experiments revealed that $\mathrm{\mu}$m-sized water-ice particles possess a specific surface energy of $(0.19\pm0.04) \, \mathrm{J \, m^{-2}}$. This is about a factor of ten higher than the specific surface energy of $\mathrm{\mu}$m-sized silica particles, which has a value of $0.02 \, \mathrm{J \, m^{-2}}$ \citep{BlumWurm2000, Gundlach2011b}.
\par
Furthermore, \citet{Aumatelletal2014} found that a specific surface energy of $0.37 \, \mathrm{J \, m^{-2}}$ provides the best fit to their experimental data. \citet{KetchamandHobbs1969} found experimentally that the specific surface energy of water ice is $0.07 \, \mathrm{J \, m^{-2}}$ in the case of ice-ice contacts at $273 \, \mathrm{K}$. In addition, \citet{DingPanetal2008} theoretically calculated values ranging from $0.19 \, \mathrm{J \, m^{-2}}$ to $0.24 \, \mathrm{J \, m^{-2}}$ for the specific surface energy of hexagonal water ice at $72\, \mathrm{K}$. Close to the melting point, \citet{Israel1992} provides a value of $0.11 \, \mathrm{J \, m^{-2}}$ for the specific surface energy of water ice. Thus, the available data suggests that the specific surface energy of water ice falls between $0.07 \, \mathrm{J \, m^{-2}}$ to $0.37 \, \mathrm{J \, m^{-2}}$.

\subsection{Temperature of the $\mathrm{\mu}$m-sized water-ice particles during the passage of the experimental setup}\label{Temperature of the ice particles before collision}
In the following, we will discuss the temperature of the $\mathrm{\mu}$m-sized water-ice particles inside the experimental setup, starting from the dispersion of $\mathrm{\mu}$m-sized water droplets at room temperature until the collision of the ice particles with the ice aggregates grown on the cold plate. Therefore, the three stages (stage I - III) in Fig. \ref{experimental_setup_skectch} are used to guide the reader through the following temperature discussion. During stage I, the  $\mathrm{\mu}$m-sized water-ice particles are produced and then cooled during their slow sedimentation. Stage II extends from the extraction of the $\mathrm{\mu}$m-sized water-ice particles from the sedimentation chamber through between the side chamber and the main chamber. The third stage (stage III) describes the passage of the $\mathrm{\mu}$m-sized water-ice particles inside the main vacuum chamber and ends with their collision with the grown ice aggregates.

\subsubsection{Temperature of the $\mathrm{\mu}$m-sized water-ice particles in stage I}
In Stage I, cooling of the produced $\mathrm{\mu}$m-sized water droplets occurs during sedimentation inside the cold sedimentation chamber whose bottom is actively cooled by liquid nitrogen. We monitored the temperature of the cold gas environment inside the sedimentation chamber during the experiments (see Tab. \ref{Tab_1}). Due to the low sedimentation speed of $\mathrm{\mu}$m-sized particles in air, the particles are in thermal equilibrium with the cold gas environment before extraction of the $\mathrm{\mu}$m-sized water-ice particles occurs\footnote{The sedimentation speed of a particle with a radius of $1.5 \, \mathrm{\mu m}$ in air is $2.9 \times 10^{-4} \, \mathrm{m \, s^{-1}}$. Hence, these particles needs $\sim 1000 \, \mathrm{s}$ to reach the height where thei extraction occurs (the distance between the droplet dispenser and the extraction is $30 \, \mathrm{cm}$). The sedimentation timescale is, thus, much longer than the cooling timescale of the particle ($1.5 \times 10^{-5}\, \mathrm{s}$).}. Thus, the temperature of the $\mathrm{\mu}$m-sized water-ice particles at the end of stage I (i.e., before extraction) equals the temperature measured by the temperature sensor inside the sedimentation chamber (see $T_{\rm sedimentation\ chamber}$ in Tab. \ref{Tab_1}). Due to the rather slow freezing process, the water-ice particles possess a hexagonal structure and are not amorphous \citep{mayerbrueggeller1982}.

\subsubsection{Temperature of the $\mathrm{\mu}$m-sized water-ice particles in stage II}\label{Temperature stage II}
In stage II, the $\mathrm{\mu}$m-sized water-ice particles are transported from the sedimentation chamber, through the side chamber, to the main chamber within a transport tube of $1\, \mathrm{mm}$ radius. The transport tube and the side chamber are located inside a liquid-nitrogen bath of $60 \, \mathrm{cm}$ length, which enables an effective cooling of this part of the experiment to $T = 77\, \mathrm{K}$. We also controlled the temperatures of the transport tube before and behind the liquid nitrogen bath ($T_7$ and $T_6$ in Fig. \ref{experimental_setup_skectch}, respectively). Due to the high heat conductivity of the material of the transport tube (aluminum), both temperatures were always below $100 \mathrm{K}$.
\par
The temperature of the $\mathrm{\mu}$m-sized water-ice particles at the end of stage II can be calculated by first deriving the gas temperature inside the transport tube. The cooling timescale of the gas inside the transport tube can be derived by solving the heat transfer equation for a gas pressure of $50\, \mathrm{mbar}$ (the typical pressure inside the first vacuum chamber; see Sect. \ref{Experimental setup}), a gas heat conductivity of $0.026 \, \mathrm{W\, m^{-1} \, K^{-1}}$ and a specific heat capacity of the gas of $1.005 \, \mathrm{kJ \, kg^{-1} \, K^{-1}}$. This calculation results in a cooling timescale of the gas inside the transport tube of $\tau_{\rm gas,\ tube} \, = \, 1.9 \times 10^{-3}\, \mathrm{s}$. For a typical velocity of $10 \, \mathrm{m \, s^{-1}}$ inside the transport tube, the gas requires $2\, \mathrm{cm}$ (e-folding length scale) to reach thermal equilibrium with the material of the transport tube. This distance is much smaller than the length of the liquid nitrogen bath, which implies that the temperature of the gas is $77 \, \mathrm{K}$ throughout the liquid nitrogen bath. The temperature of the transport tube behind the liquid nitrogen bath (at the end of stage II) has been measured to be always below $100 \, \mathrm{K}$, thus, the gas temperature at the end of this stage is between $77 \, \mathrm{K}$ and $100 \, \mathrm{K}$.
\par
In order to derive the ice-particle temperature at the end of stage II, the thermal coupling between the $\mathrm{\mu}$m-sized water-ice particles and the gas inside the transport tube has to be calculated. Therefore, we used the heat exchange equation of a particle-laden gas flow derived by \citet[][his Eq. 16]{Jones1991} with a specific heat capacity of water ice of $2100 \, \mathrm{J \, kg^{-1} \, K^{-1}}$, a mass density of water ice of $930 \, \mathrm{kg \, m^{-1}}$, a mean radius of the ice particles of $1.5\, \mathrm{\mu m}$, a Nusselt number of unity, an emissivity of unity, and an initial temperature of the ice particles of $100 \, \mathrm{K}$ (however, the calculations have shown that the cooling of the $\mathrm{\mu}$m-sized water-ice particles is dominated by heat conduction, which implies that the initial temperature of the ice particles has a negligible influence on the derived timescale) to derive the cooling timescale of the $\mathrm{\mu}$m-sized water-ice particles inside the transport tube of $\tau_{\rm ice,\ tube} \, = \, 1.5 \times 10^{-4}\, \mathrm{s}$.
\par
With a typical velocity of the gas and the $\mathrm{\mu}$m-sized water-ice particles inside the transport tube of $10 \, \mathrm{m \, s^{-1}}$ and the above-derived cooling timescale of the $\mathrm{\mu}$m-sized water-ice particles, we get a typical e-folding length of the $\mathrm{\mu}$m-sized water-ice particles to follow temperature changes of the gas of only $2\, \mathrm{mm}$. This length is much smaller than the length of the cooled transport tube, which implies that the temperature of the $\mathrm{\mu}$m-sized water-ice particles at the end of this stage equals the gas temperature (ranging from $77 \, \mathrm{K}$ to $100 \, \mathrm{K}$).

\subsubsection{Temperature of the $\mathrm{\mu}$m-sized water-ice particles in stage III}
In order to derive the temperature evolution of the $\mathrm{\mu}$m-sized water-ice particles inside the main vacuum chamber, stage III is divided into two parts. The first part describes the passage of the $\mathrm{\mu}$m-sized water-ice particles through the transport tube inside the main chamber until the end of the transport tube (exit nozzle) is reached. The transport tube inside the main chamber is surrounded by a cooling system (see copper shield in Fig. \ref{experimental_setup_skectch} and Fig \ref{cooling_system} for an image of the cooling system) to provide low temperatures during the experiments. The second part is the free flight of the $\mathrm{\mu}$m-sized water-ice particles inside the main chamber between the exit nozzle and the ice aggregate where the collisions occur. This second part is also fully surrounded by the cooling system, a hollow tube with an inner radius of $2\, \mathrm{cm}$.
\par
The derivation of the gas and the ice-particle temperature in the first part of stage III is similar to the temperature discussion performed for stage II. The transport tube is actively cooled by the cooling system over a length of $15 \, \mathrm{cm}$. We permanently measured the temperature of the transport tube at two positions during the experimental runs ($T_3$ and $T_5$ in Fig. \ref{experimental_setup_skectch}, respectively) to ensure temperature stability. The thermal coupling distances (e-folding distances) of the gas and the $\mathrm{\mu}$m-sized water-ice particles are again $2\, \mathrm{cm}$ and $2\, \mathrm{mm}$, respectively (see Sect. \label{Temperature stage II}). These distances are much smaller than the cooled length of the transport tube. Hence, the gas and the $\mathrm{\mu}$m-sized water-ice particles are in thermal equilibrium with the end of the transport tube (exit nozzle). The temperature of the nozzle was measured ($T_5$ in Fig. \ref{experimental_setup_skectch}) and, thus, the temperature of the $\mathrm{\mu}$m-sized water-ice particles at this position is known (see $T_{\rm nozzle}$ in Tab. \ref{Tab_1}).
\par
During the free flight of the $\mathrm{\mu}$m-sized water-ice particles between the exit nozzle and the ice aggregate (second part of stage III), cooling occurs inside the cylindrically shaped cooling system. The temperature calculation of gas and ice particles is similar to the first part of this stage. First, the cooling timescale for the gas in the main chamber is calculated. In this case, a pressure of $1\,\mathrm{mbar}$ and a radius of the cooling system of $2\, \mathrm{cm}$ are used. The derived cooling é-folding timescale of the gas inside the cooling system is $\tau_{\rm gas,\ cooling\ system} \, = \, 1.5 \times 10^{-2}\, \mathrm{s}$, which implies that the gas requires an e-folding distance of $15\, \mathrm{cm}$ to reach the temperature of the cooling system (again assuming a typical velocity of $10 \, \mathrm{m \, s^{-1}}$). Because the gas cools roughly linearly for distances much shorter than the cooling length, the gas temperature at the position of the cold plate can be derived. Here, we exemplarily calculate the temperature of the gas in case of the experimental run number 1 (see Tab. \ref{Tab_1}) for which the difference between the temperature of the nozzle and the temperature of the cooling system was maximal ($\sim 130 \, \mathrm{K}$). In this case, the gas temperature has decreased by $17 \,\mathrm{K}$ between the exit nozzle and the position of the cold target (the distance between the nozzle and the cold plate is $2\, \mathrm{cm}$). This is the maximum cooling the gas has experienced inside the cooling system during the six different experimental runs.
\par
With the knowledge of the gas temperature at the position of the cold plate, the temperature of the $\mathrm{\mu}$m-sized water-ice particles can be derived. Therefore, the cooling timescale of the $\mathrm{\mu}$m-sized water-ice particles is estimated by again using the exchange of a particle laden gas flow derived by \citet[][his Eq. 16; see Sect. \ref{Temperature stage II}]{Jones1991}, but in this case for a pressure of $1\,\mathrm{mbar}$. The thus-calculated cooling timescale of the $\mathrm{\mu}$m-sized water-ice particles inside the cooling system is $\tau_{\rm ice,\ cooling\ system} \, = \, 7.5 \times 10^{-3}\, \mathrm{s}$. This implies that the $\mathrm{\mu}$m-sized water-ice particles require an e-folding distance of $8 \, \mathrm{cm}$ (for a typical velocity of $10 \, \mathrm{m \, s^{-1}}$) to reach the gas temperature. Because of this cooling distance, the $\mathrm{\mu}$m-sized water-ice particles experienced only 1/4 of the gas temperature decrease at the position of the ice aggregate ($2 \, \mathrm{cm}$ behind the nozzle) if a linear temperature decrease with distance is again assumed. In the case of experimental run number 1 (the experimental run with maximum cooling of the gas inside the cooling system), the $\mathrm{\mu}$m-sized water-ice particles have cooled by $4\, \mathrm{K}$ when they reach the position of the cold plate. The temperature decrease of the $\mathrm{\mu}$m-sized water-ice particles during the other experimental runs was always less than $4\, \mathrm{K}$, because the experimental run number 1 enabled for maximal cooling due to the highest temperature difference between the exit nozzle and the cooling system (see Tab. \ref{Tab_1}). Thus, the temperature decrease of the $\mathrm{\mu}$m-sized water-ice particles between the nozzle and the cold plate is less than the variation of the temperature of the nozzle during the experiments. This implies that the temperature of the $\mathrm{\mu}$m-sized water-ice particles can be well approximated by the measured temperature of the exit nozzle.

\subsection{Temperature of the grown ice aggregates composed of the $\mathrm{\mu}$m-sized water-ice particles}\label{Temperature of the grown ice aggregates}
The grown ice aggregates possess porous structures composed of $\mathrm{\mu}$m-sized water-ice particles. We assume that the grown ice aggregates obtain a porosity of approximately 0.5. The heat conductivity of porous aggregates in vacuum is generally very low \citep[in the order of $10^{-2} \, \mathrm{W \, m^{-1} \, K^{-1}}$; ][]{Krause2011, Gundlach2012}. Thus, the temperature of the region of the ice aggregates where the collisions with the impinging $\mathrm{\mu}$m-sized water-ice particles occur (see Fig. \ref{growth_1}) is in thermal equilibrium with the ambient temperature provided by the copper cylinder (I in Fig. \ref{cooling_system}) of the cooling system. Additionally, the temperature of the cold plate (II in Fig. \ref{cooling_system}) is equal to the temperature of the copper cylinder during the experimental runs, which implies that the temperature gradient inside the sample is very small. Thus, the temperature of the grown ice aggregates is in equilibrium with the temperature of the cooling system (see Tab. \ref{Tab_1}).


\section{Experimental results}\label{Experimental results}

\subsection{Growth and erosion of ice aggregates composed of $\mathrm{\mu}$m-sized water-ice particles}\label{Growth and erosion of ice aggregates}
\begin{figure*}[t!]
\centering
\includegraphics[angle=0,width=1\textwidth]{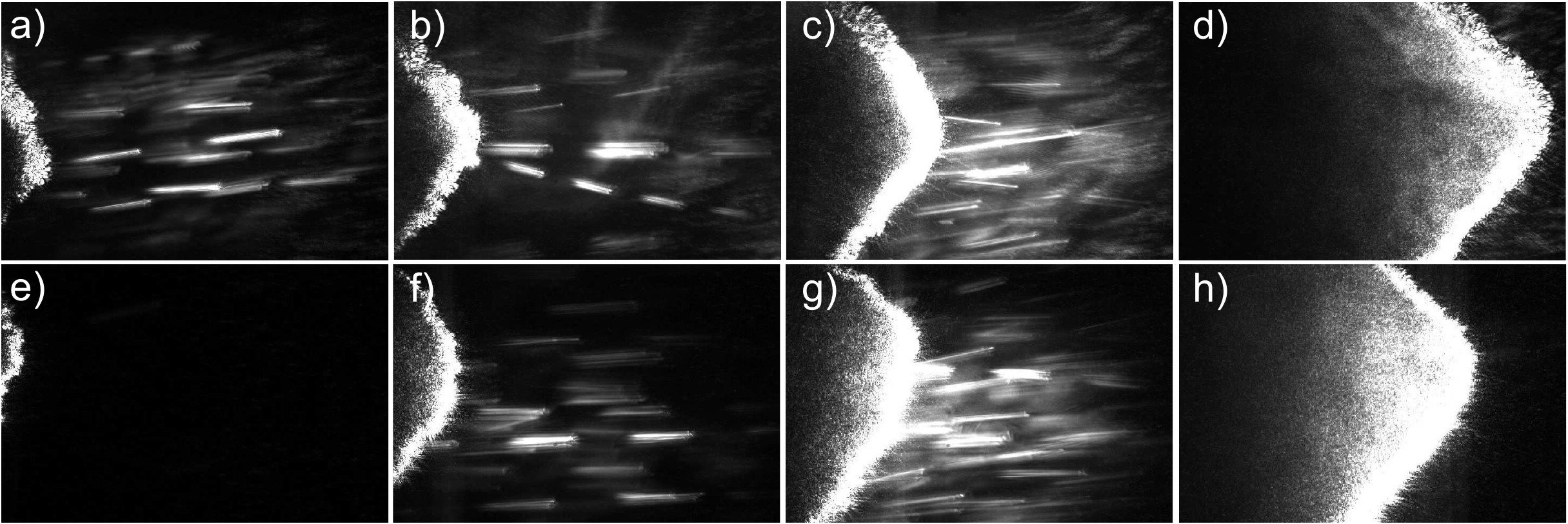}
\caption{Two image sequences of the growth of ice aggregates (images a-d and images e-h) by sticking of impinging $\mathrm{\mu}$m-sized water-ice particles. The images were taken with the long-distance microscope and the aggregates were illuminated by the pulsed laser beam. In some images, impinging $\mathrm{\mu}$m-sized water-ice particles were observed (dashed lines). The velocities of these water-ice particles were ranging from $4.7\,\mathrm{m \, s^{-1}}$ to $13.9\,\mathrm{m \, s^{-1}}$. The time differences between the first and the last image of the image sequences are $22 \, \mathrm{ms}$ (image d - image a) and $28 \, \mathrm{ms}$ (image h - image e), respectively. Note that also bouncing of some $\mathrm{\mu}$m-sized water-ice particles occurred during the growth of the ice aggregates (e.g., in image b). Both ice aggregates were grown during the experimental run number 6 (see Tab. \ref{Tab_1}). The dimensions of the images are $612\, \mathrm{\mu m} \times 408\, \mathrm{\mu m}$.}
\label{growth_1}
\end{figure*}

\begin{figure*}[t!]
\centering
\includegraphics[angle=0,width=1\textwidth]{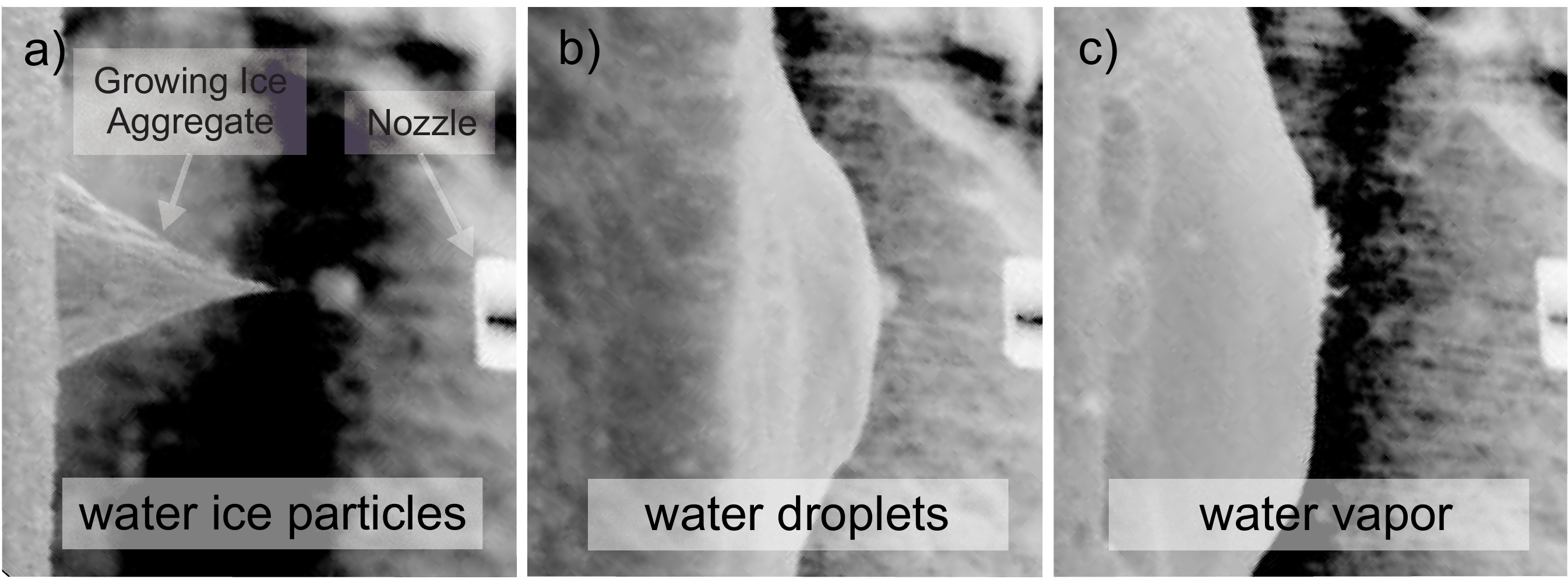}
\caption{Grown ice targets observed through the top window of the vacuum chamber. Image a shows an ice aggregate grown by the impact of  $\mathrm{\mu}$m-sized water-ice particles. Images b and c show icy deposits on the cooled target grown by the impact of uncooled water droplets (b) and water vapor deposition (c), respectively. The colors of the images are inverted in order to enhance the visibility of the ice targets. Note that in the case of the water vapor deposition (c), also frost is visible on top of the solid ice target. The dimensions of the images are $2.3\, \mathrm{cm} \times 2.5\, \mathrm{cm}$.}
\label{growth_2}
\end{figure*}

With the experimental setup described above (see Sect. \ref{Experimental setup}), we performed impact experiments with the $\mathrm{\mu}$m-sized water-ice particles onto the cryogenically cooled target. As explained in Sects. \ref{Temperature of the grown ice aggregates} and \ref{Temperature of the ice particles before collision}, the temperature of the impinging water-ice particles is approximately given by the temperature of the nozzle and the temperature of the target water-ice agglomerate is that of the cooling system. Both temperatures are presented in Tab. \ref{Tab_1} for the six experimental runs. As long as the impact velocities were sufficiently small, porous ice aggregates formed on the target by direct sticking of the impinging $\mathrm{\mu}$m-sized water-ice particles. The growing ice aggregates were observed by the long-distance microscope and the laser setup (see Fig. \ref{experimental_setup}). Fig. \ref{growth_1} shows two examples of the growth of ice aggregates. The growing ice aggregates were also observed through the top window of the vacuum chamber utilizing a standard CCD camera (see Fig. \ref{growth_2}a). For comparison, we also collided uncooled water micro-droplets with the cryogenically cooled target (see Fig. \ref{growth_2}b) and sprayed water vapor onto the cooled target (see Fig. \ref{growth_2}c). We performed these different experiments in order to demonstrate the influence of the projectile state (solid, liquid, vapor) on the visual appearance of the growing ice targets.
\par
\begin{figure*}[t!]
\centering
\includegraphics[angle=0,width=1\textwidth]{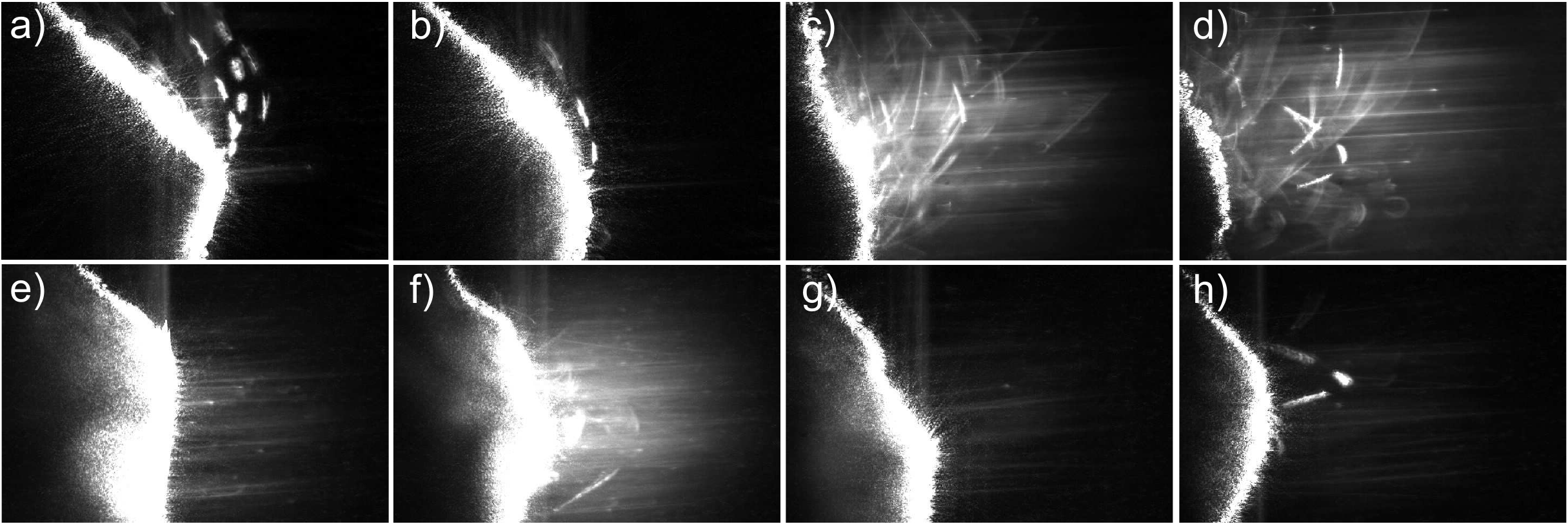}
\caption{The two image sequences show the erosion of two different ice aggregates (images a-d and images e-h) composed of $\mathrm{\mu}$m-sized water-ice particles when the impinging water-ice particles impinge above the erosion threshold velocity. The images were taken with the long-distance microscope and illumination is done by a pulsed laser beam. The velocities of the impinging $\mathrm{\mu}$m-sized water-ice particles (lines) are ranging between $17.7\,\mathrm{m \, s^{-1}}$ and $49.3\,\mathrm{m \, s^{-1}}$. The time differences between the first and the last image of the image sequences are $12 \, \mathrm{ms}$ (image d - image a) and $8 \, \mathrm{ms}$ (image h - image e), respectively. The two image sequences were taken during the experimental runs number 5 and 6, respectively (see Tab. \ref{Tab_1}). The dimensions of the images are $612\, \mathrm{\mu m} \times 408\, \mathrm{\mu m}$.}
\label{erosion_1}
\end{figure*}

At higher velocities, the ice aggregates get eroded by the impinging $\mathrm{\mu}$m-sized water-ice particles. Fig. \ref{erosion_1} exemplarily shows the erosion of two different ice aggregates. For velocities higher than the sticking threshold and lower than the erosion threshold, bouncing of the $\mathrm{\mu}$m-sized water-ice particles occurred. In this case, no growth and no erosion of the ice aggregates was observed (see Sect. \ref{The sticking and the erosion threshold of micrometer-sized water-ice particles}).


\subsection{Derivation of the sticking and erosion thresholds of $\mathrm{\mu}$m-sized water-ice particles}\label{The sticking and the erosion threshold of micrometer-sized water-ice particles}
In order to measure the threshold velocities for sticking and erosion of the $\mathrm{\mu}$m-sized water-ice particles, the growth and erosion behavior of the ice aggregates was observed using the long-distance microscope and the laser-illumination setup (see Figs. \ref{growth_1} and \ref{erosion_1}). For each growth (erosion) sequence, the growth (erosion) rate of the ice aggregate (measured in units of $\rm \mu m$ growth length per second exposure of the target to the impinging projectile flux) and the corresponding velocities of the impinging $\mathrm{\mu}$m-sized water-ice particles was measured. Additionally, sequences in which no growth and no erosion occurred were identified as dominated by bouncing of the water-ice particles. Also in these sequences, the velocities of the impinging water-ice particles were measured and the corresponding growth rate of the ice aggregate was set to zero.
\par
\begin{figure*}[p!]
\includegraphics[angle=90,width=0.49\textwidth]{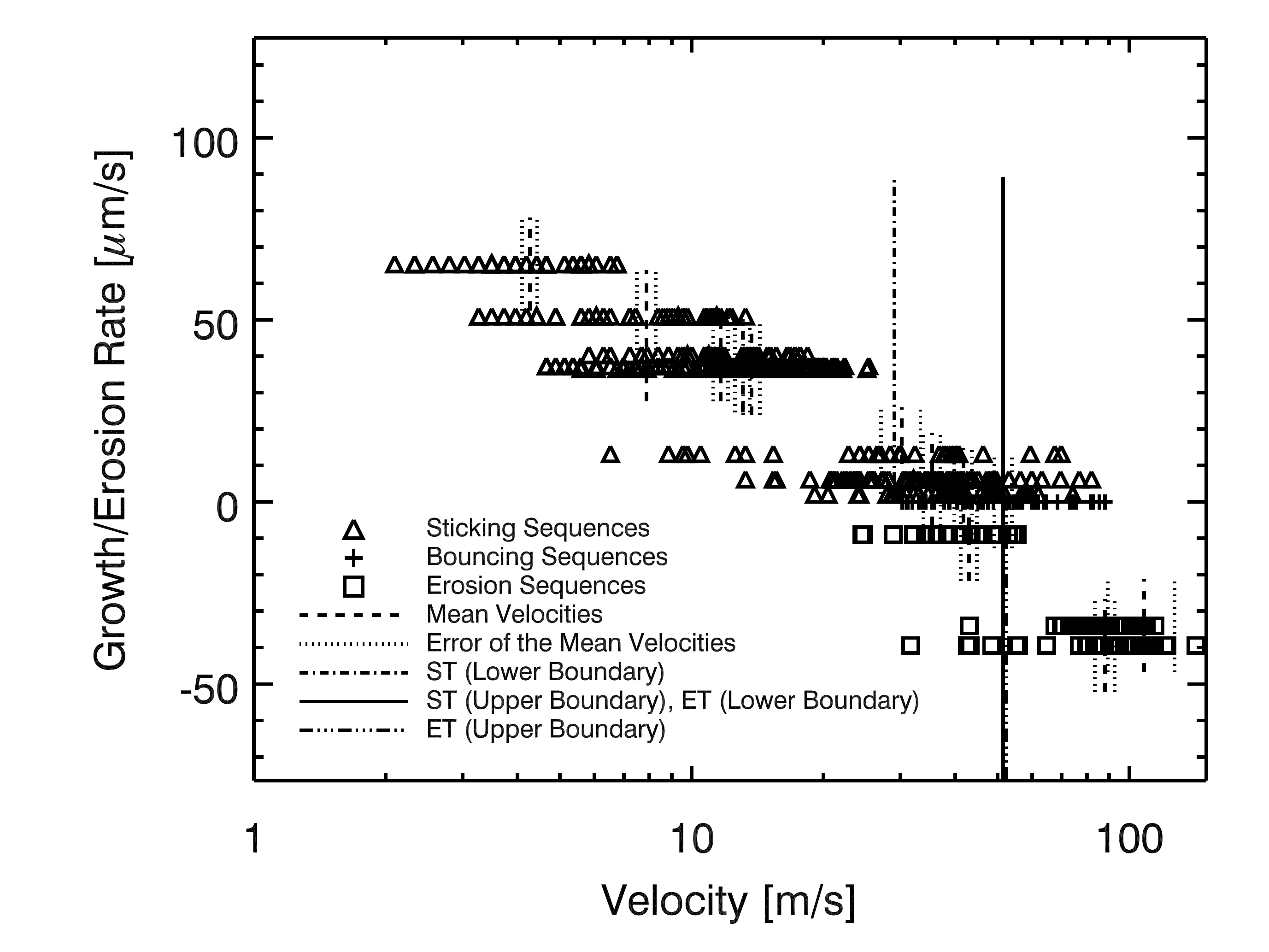}
\put(-176,147){\Large{\bf{a)}}}
\put(-160,147){Exp. 1}
\put(-68,147){$(255\pm5)\,\mathrm{K}$}
\put(54.5,147){\Large{\bf{b)}}}
\put(71,147){Exp. 2}
\put(162,147){$(239\pm4)\,\mathrm{K}$}
\includegraphics[angle=90,width=0.49\textwidth]{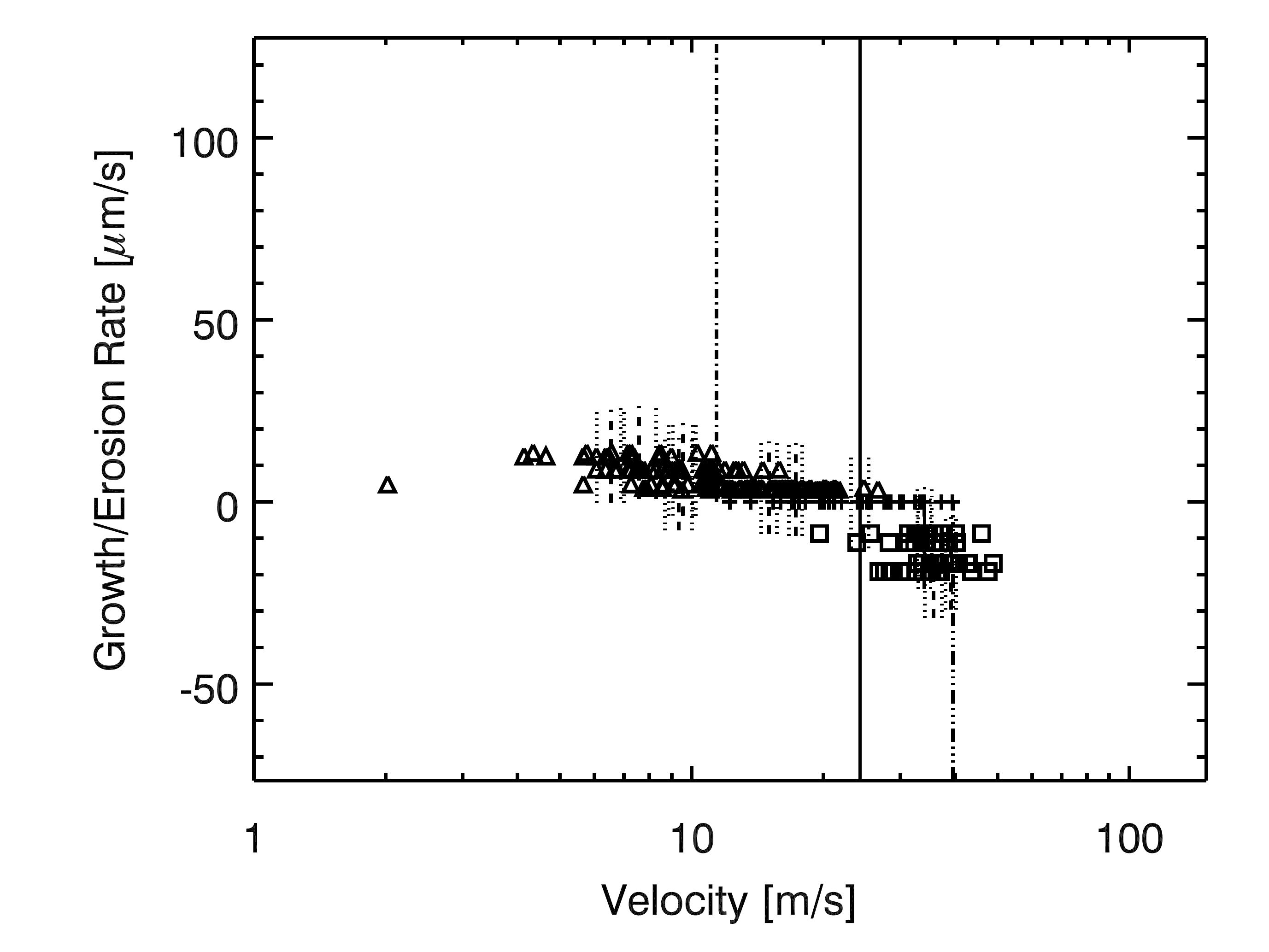}
\includegraphics[angle=90,width=0.49\textwidth]{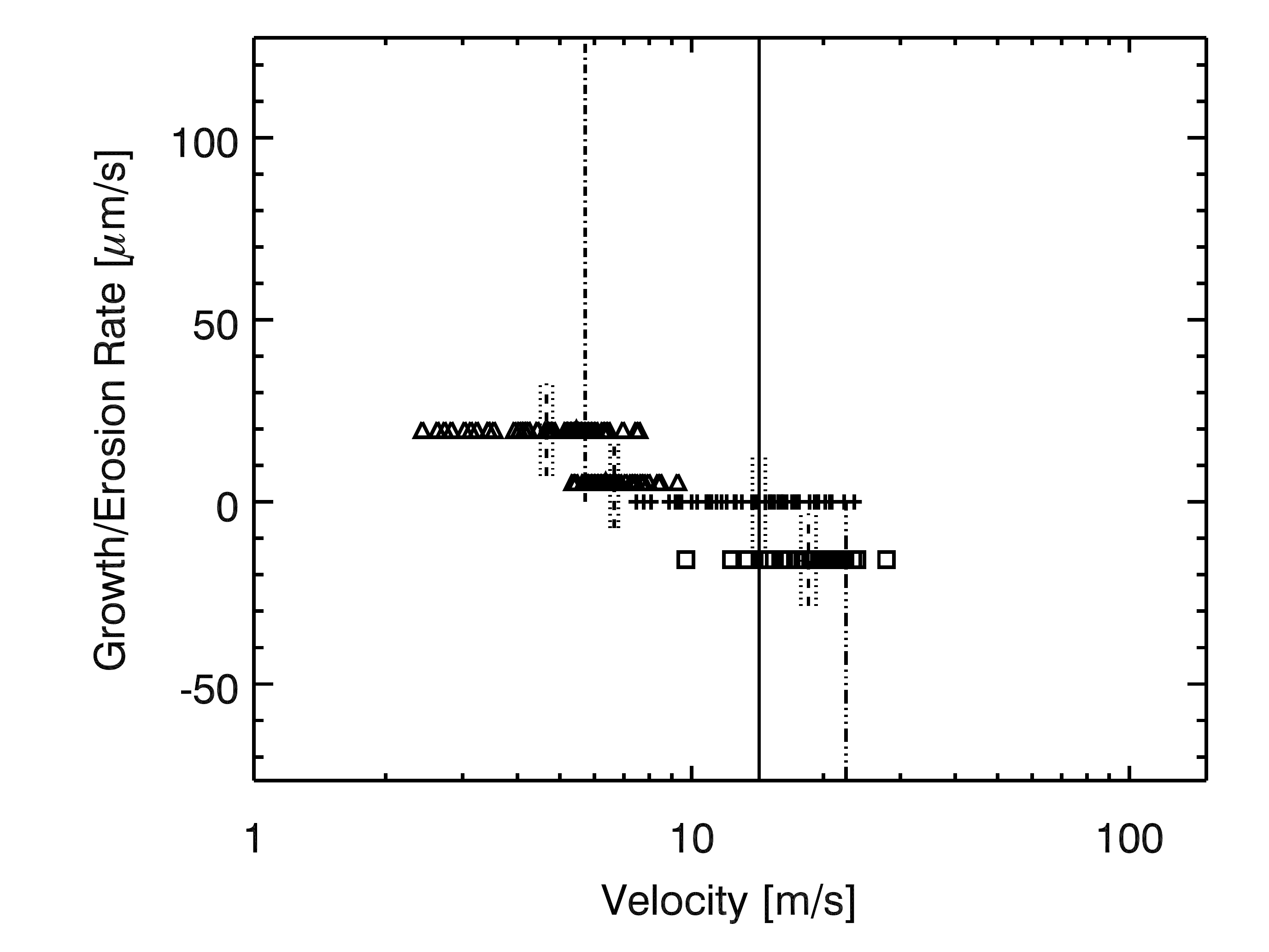}
\put(-176,147){\Large{\bf{c)}}}
\put(-160,147){Exp. 3}
\put(-68,147){$(203\pm5)\,\mathrm{K}$}
\put(54.5,147){\Large{\bf{d)}}}
\put(71,147){Exp. 4}
\put(157,147){$(165\pm15)\,\mathrm{K}$}
\includegraphics[angle=90,width=0.49\textwidth]{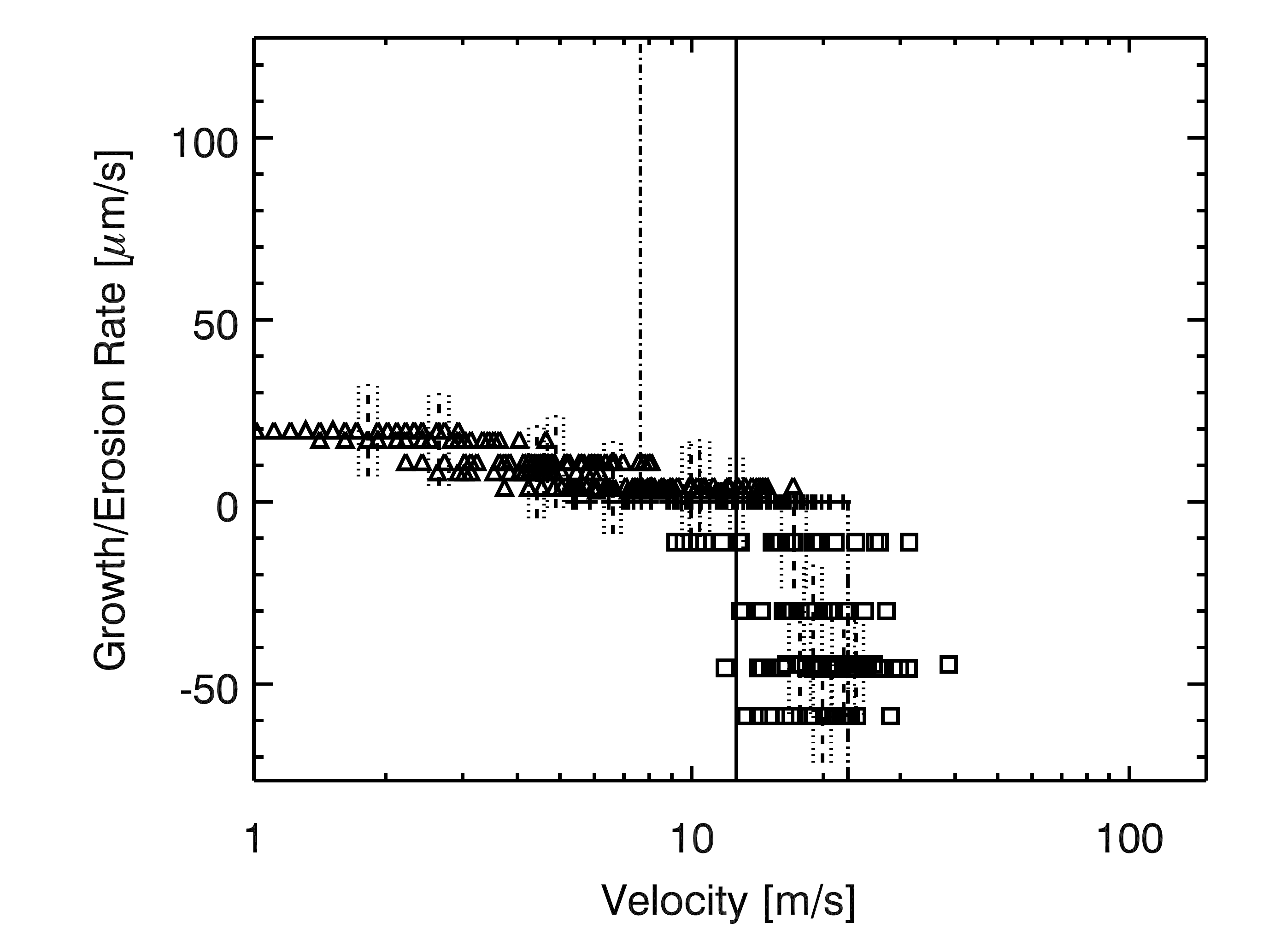}
\includegraphics[angle=90,width=0.49\textwidth]{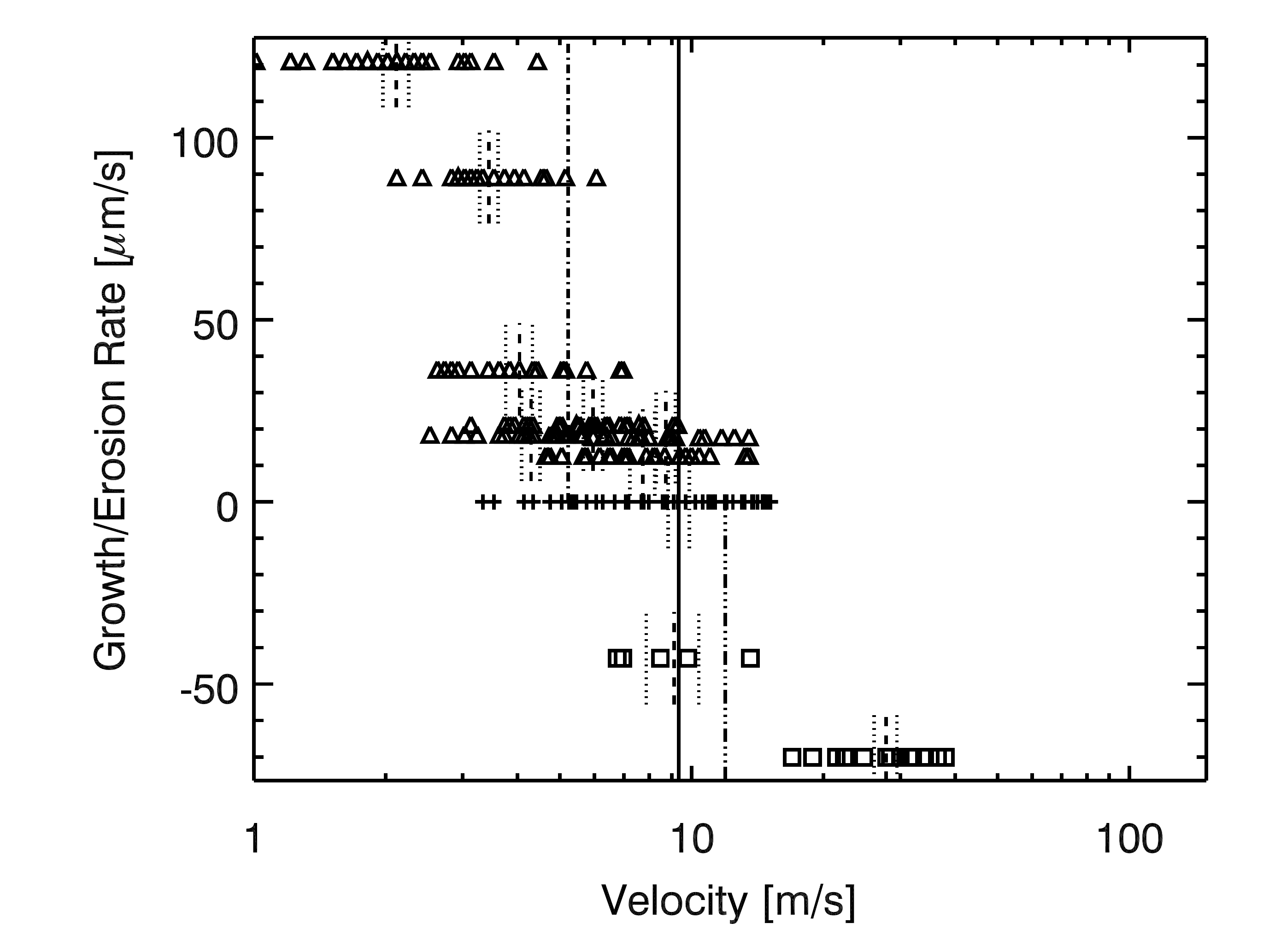}
\put(-176,40){\Large{\bf{f)}}}
\put(-160,40){Exp. 5}
\put(-68,147){$(143\pm5)\,\mathrm{K}$}
\put(54.5,147){\Large{\bf{g)}}}
\put(71,147){Exp. 6}
\put(162,147){$(117\pm3)\,\mathrm{K}$}
\includegraphics[angle=90,width=0.49\textwidth]{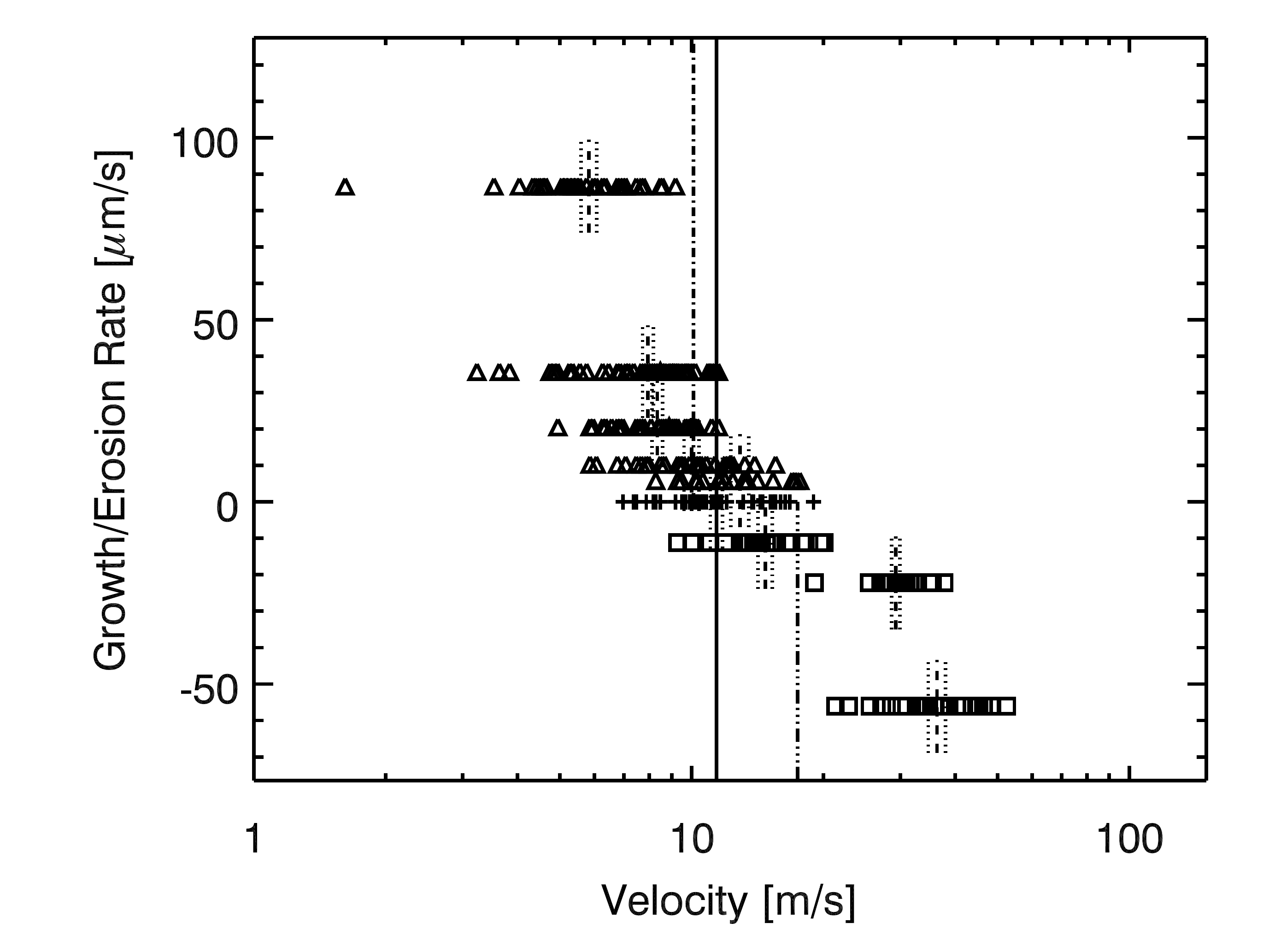}
\caption{Observed rates for growth (triangles), bouncing (pluses), and erosion (squares) during the different experimental runs. Each panel shows an individual experimental run, performed at different temperatures. The temperatures of the $\mathrm{\mu}$m-sized water-ice particles at the position of the ice aggregates are also shown (see Sect. \ref{Temperature of the ice particles before collision} and Tab. \ref{Tab_1}). In case of a positive growth rate, the ice aggregates grew in size with time (growth; see Fig. \ref{growth_1}). A growth rate of zero denotes that no growth and no erosion of the ice aggregates was observed (bouncing). Negative values of the growth rate means that erosion of the ice aggregates occurred (erosion; see Fig. \ref{erosion_1}). For each sequence, the mean velocity (dashed lines) and the standard error of the mean velocity (dotted lines) are shown. The lower and upper boundaries of the sticking threshold (ST) are shown by the dash-dotted and solid vertical lines, respectively. Similarly, the lower and upper boundaries of the erosion threshold (ET) are denoted by the solid and the dash-triple-dotted lines.}
\label{sticking_comp}
\end{figure*}

The results of this analysis are summarized in Fig. \ref{sticking_comp} for the different experimental runs, which were performed at different temperature (see Tab. \ref{Tab_1} and Sects. \ref{Temperature of the grown ice aggregates} and \ref{Temperature of the ice particles before collision}). Each panel of Fig. \ref{sticking_comp} shows an individual experimental run during which the temperature of the nozzle and, therewith, the temperature of the impinging $\mathrm{\mu}$m-sized water-ice particles were kept constant (see Tab. \ref{Tab_1}). In Fig. \ref{sticking_comp} the velocities obtained during the growth, bouncing, and erosion sequences are denoted by the triangles, pluses, and squares, respectively. For each of these sequences, the mean velocity (dashed lines) of the $\mathrm{\mu}$m-sized water-ice particles and the standard error $\sigma$ of the mean velocity (dotted lines) are shown. It is important to mention that the shown growth rates are absolute values and are not normalized, because it was not possible to determine the number of impinging particles per unit area and time during the experiments.

\par
In the following, we will explain how the sticking threshold of the $\mathrm{\mu}$m-sized water-ice particles was derived from the measured data. In order to be independent from model assumptions, we decided to determine a lower and an upper boundary to the sticking threshold for each temperature, which provides a maximum range within which the sticking threshold velocity can be found.
\par
The lower boundary of the sticking threshold was derived from the lowest growth rates observed during the different experimental runs. We assume that in this case, all collisions with velocities lower than the mean velocities minus the corresponding 1$\sigma$-deviations of the mean velocities led to a growth of the ice aggregates. These lower boundaries of the sticking threshold are shown as dash-dotted lines in Fig. \ref{sticking_comp}. The upper boundary of the sticking threshold was calculated by the assumption that in the case of bouncing, all impinging particles with velocities lower than the mean velocity caused growth of the ice aggregate whereas all water-ice particles impinging above the mean velocity cause erosion so that both processes just cancel. Thus, the upper boundaries of the sticking threshold are given by the mean values of the velocities measured during the bouncing sequences (solid lines in Fig. \ref{sticking_comp}).
\par
\begin{figure}[t!]
\centering
\includegraphics[angle=90,width=1\columnwidth]{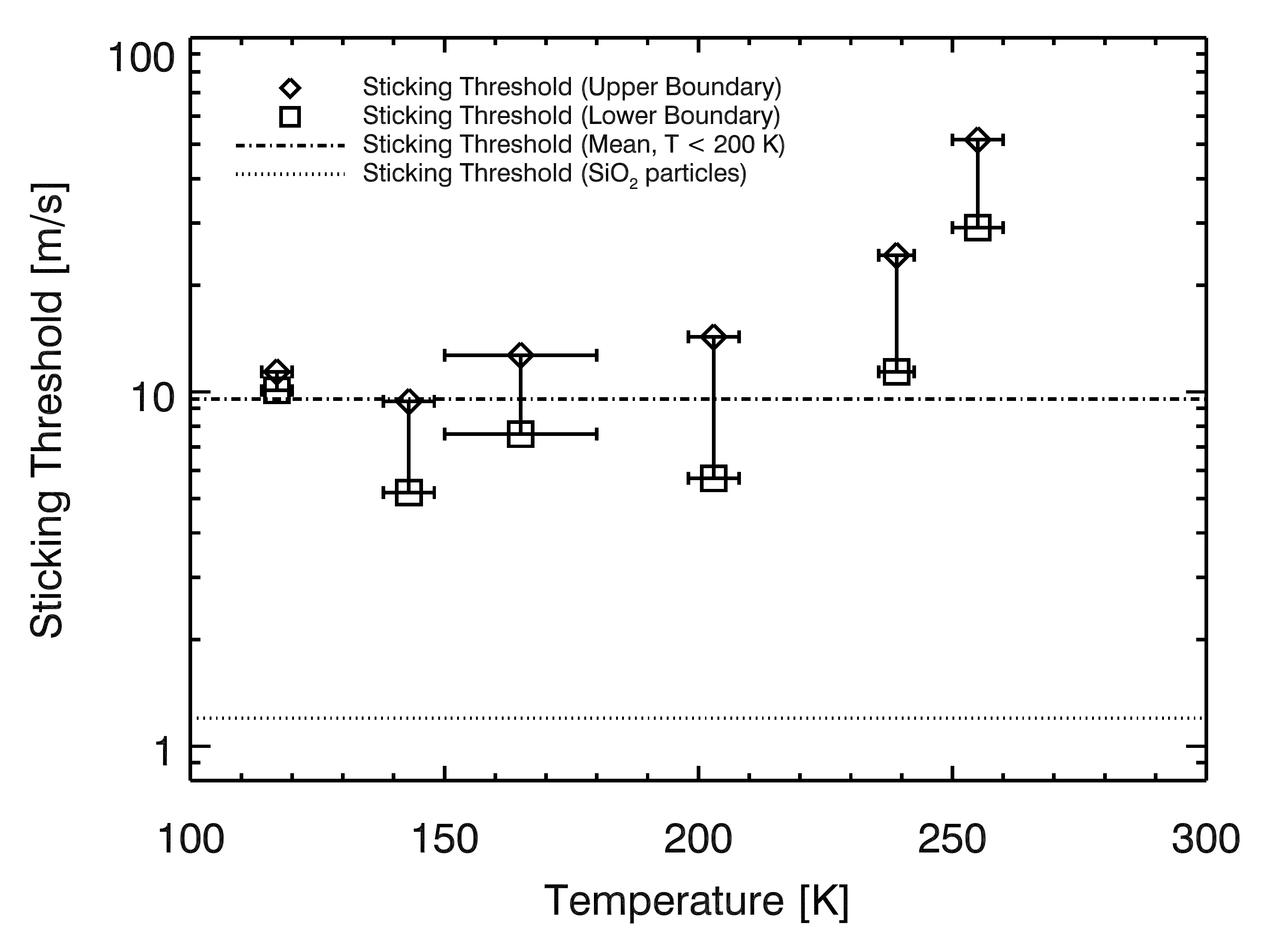}
\put(-37,98){$\mathrm{H_2O}$}
\put(-37,43.5){$\mathrm{SiO_2}$}
\caption{Temperature dependent threshold velocity for sticking of the $\mathrm{\mu}$m-sized water-ice particles. The lower and the upper boundaries of the sticking threshold are shown by the squares and the diamonds, respectively. For temperatures below $\sim 210 \, \mathrm{K}$, the sticking threshold is within the measurement uncertainties constant at $9.6 \, \mathrm{m \, s^{-1}}$ (dash-dotted line). For comparison, the sticking threshold for spherical silica particles with radii of $0.6 \, \mathrm{\mu m}$ is $1.2 \, \mathrm{m\, s^{-1}}$ and is shown by the dotted line \citep{Poppe2000}.}
\label{Sticking_Threshold_Temperature_1}
\end{figure}

Fig. \ref{Sticking_Threshold_Temperature_1} shows the derived lower and upper boundaries of the sticking threshold (squares and diamonds, respectively) as a function of the temperature of the impinging $\mathrm{\mu}$m-sized water-ice particles. For temperatures below $\sim 210 \, \mathrm{K}$, the sticking threshold of the $\mathrm{\mu}$m-sized water-ice particles is within the measurement uncertainties constant at $9.6 \, \mathrm{m \, s^{-1}}$ (dashed-dotted line; this value is derived by calculating the arithmetic mean of the lower and upper boundaries obtained for each of the four experimental runs performed at temperatures below $\sim 210 \, \mathrm{K}$). For higher temperatures, the sticking threshold increases with increasing temperature to values of $\sim 18\, \mathrm{m \, s^{-1}}$ for $T = (239 \pm 4) \, \mathrm{K}$ and $\sim 40\, \mathrm{m \, s^{-1}}$ for $T = (255 \pm 5) \, \mathrm{K}$.
\par
The measured sticking threshold of the $\mathrm{\mu}$m-sized water-ice particles at low temperatures is approximately ten times higher than the sticking threshold derived for $\mathrm{\mu}$m-sized silica particles. The latter was measured by \citet{Poppe2000} who found a sticking threshold of $1.2 \, \mathrm{m\, s^{-1}}$ for silica spheres with radii of $0.6\, \mathrm{\mu m}$ (dotted line in Fig. \ref{Sticking_Threshold_Temperature_1}). The higher sticking threshold of the $\mathrm{\mu}$m-sized water-ice particles in comparison to the $\mathrm{\mu}$m-sized silica particles is mainly due to the difference of the specific surface energies of the two materials.
\par
Between $ \sim 210 \, \mathrm{K}$ and $\sim230 \, \mathrm{K}$, the sticking threshold changes from a temperature-independent value to an increasing function with increasing temperature. This transition is observed in the same temperature region where \citet{Higaetal1998} proposed a change of the brittle states of water ice. Whether a transition of the brittle states of water ice can increase the stickiness of water ice at higher temperatures is, however, speculative.
\par
In addition to the derivation of the sticking threshold, we also investigated the erosion threshold of the $\mathrm{\mu}$m-sized water-ice particles when impinging an ice-particle target (see solid and dash-triple-dotted lines in Fig. \ref{sticking_comp}). To derive the lower and the upper boundary of the erosion threshold, the same algorithm as in the case of the sticking threshold was used. This means that the lower boundary of the erosion threshold is given by the mean velocity of the bouncing sequence (which is, thus, equal to the upper threshold for sticking). The upper boundary of the erosion threshold was derived from the lowest erosion rates observed during the different experimental runs. In these cases, we assume that all collisions with velocities higher than the mean velocities plus the corresponding 1$\sigma$-deviations led to erosion of the ice aggregates.
\par
\begin{figure}[t!]
\centering
\includegraphics[angle=90,width=1\columnwidth]{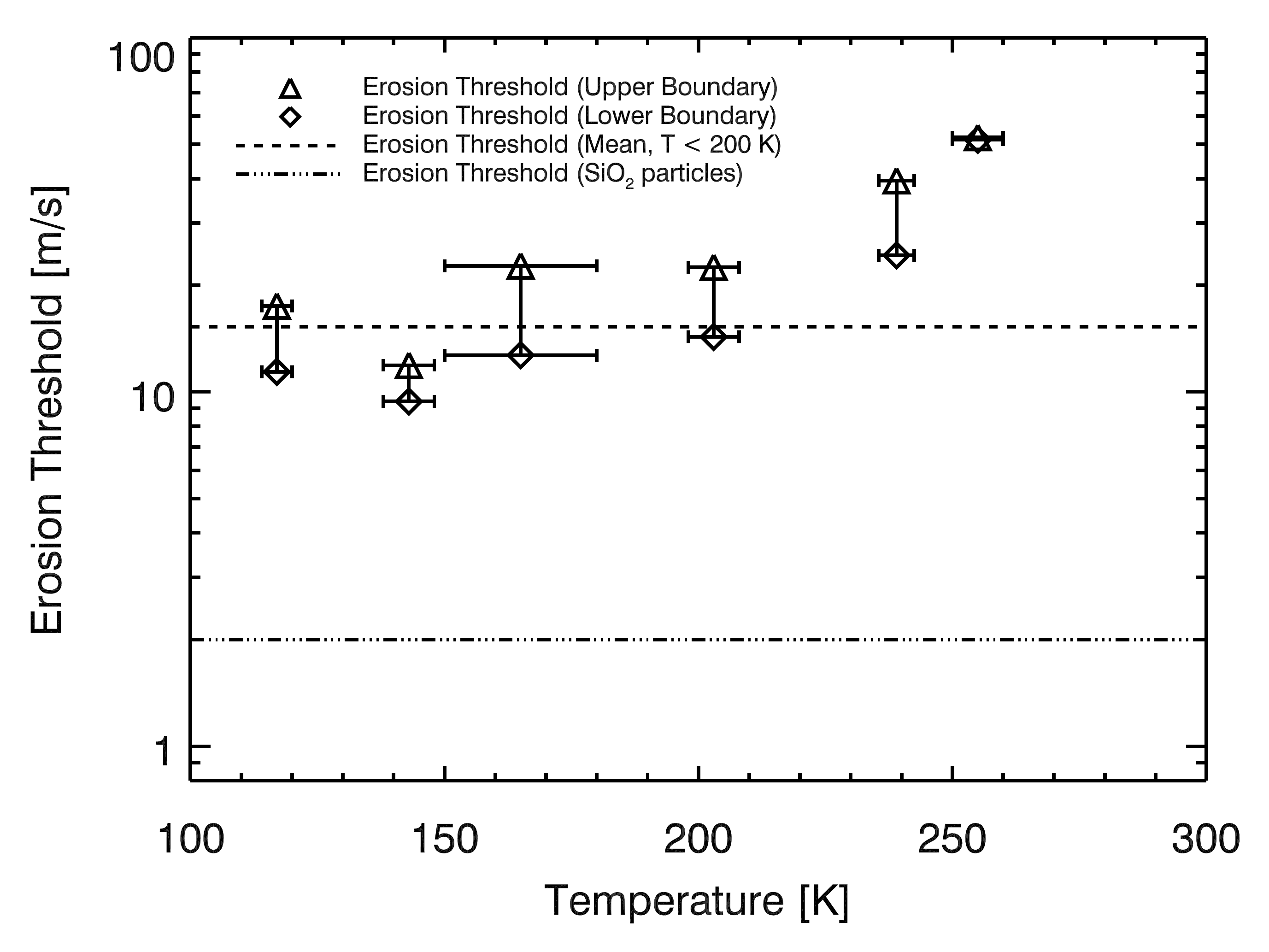}
\put(-37,110.6){$\mathrm{H_2O}$}
\put(-37,57){$\mathrm{SiO_2}$}
\caption{Temperature dependent threshold velocity for erosion of the $\mathrm{\mu}$m-sized water-ice particles. The diamonds and triangles show the lower and upper boundaries of the erosion threshold, respectively. The erosion threshold is within the measurement uncertainties constant at $15.3 \, \mathrm{m \, s^{-1}}$ for temperatures below $\sim 210 \, \mathrm{K}$ (dashed line). For comparison, the erosion threshold of spherical silica particles with radii of $0.75 \, \mathrm{\mu m}$ and for unpassivated target aggregates was derived by \citet{Schraepler2011} to be $2 \, \mathrm{m\, s^{-1}}$ and is shown by the dashed-triple-dotted line.}
\label{erosion_threshold_temperature}
\end{figure}
In Fig. \ref{erosion_threshold_temperature}, the lower and upper boundaries for the erosion threshold are shown by the diamonds and triangles as a function of temperature of the impinging water-ice particles. The erosion threshold at low temperatures (below $\sim 210 \, \mathrm{K}$) is within the uncertainties of our measurements constant at $15.3 \, \mathrm{m \, s^{-1}}$ and increases to $\sim 50 \, \mathrm{m \, s ^{-1}}$ for higher temperatures. For comparison, the dashed line indicates the erosion threshold of $\sim 2 \, \mathrm{m \, s^{-1}}$ derived for spherical silica particles with radii of $0.75 \, \mathrm{\mu m}$ and for unpassivated aggregate targets \citep{Schraepler2011}. A very similar value of the erosion threshold was also found by \citet{Seizingeretal2013} who used computer simulations in order to determine the erosion rate of silica aggregates. \citet{Schraepler2011} noted that passivation of dust aggregates occurs when more than $2 \, \mathrm{mg \, cm^{-2}}$ of dust collides with an aggregate. We used this critical value to check whether the water ice aggregates are being passivated by the impinging water-ice particles before the erosion of the ice aggregates is started. In our experiments, the ice aggregates are grown with a maximum speed of $\sim 100 \, \mathrm{\mu m \ s^{-1}}$ and we assume that the ice aggregates posses a porosity of 0.5. With the knowledge of the growth speed and the assumed porosity it is possible to derive the required time until passivation of the ice aggregates occurs (i.e., when the critical value of $2 \, \mathrm{mg \, cm^{-2}}$ is reached), which reads $t \geq 400 \, \mathrm{s}$. In our experiments, the time interval between growth and erosion of the ice aggregates was always less than $60 \, \mathrm{s}$, which implies that the ice aggregates are not passivated before erosion.
\par
A comparison between the mean sticking threshold (see Fig. \ref{Sticking_Threshold_Temperature_1}) and the mean erosion threshold (see Fig. \ref{erosion_threshold_temperature}) for the two materials shows that the ratio of the threshold velocities for erosion and sticking is $\sim 1.5$ in the case of the $\mathrm{\mu}$m-sized water-ice particles and $\sim 2$ in the case of the $\mathrm{\mu}$m-sized silica particles. These values are in reasonable agreement with the calculation performed by \citet{DominikTielens1997}. They showed that the critical energy for sticking (their Eq. 6) is $4.8$ times smaller than the energy required to break an existing contact between the particles (their Eq. 8). In terms of velocity (their Eq. 9), this means that the sticking threshold is $\sqrt{4.8} \, = \,  2.2$ times lower than the erosion threshold. However, we would like to remind the reader that the sticking threshold estimated from our experiments is derived for particle-aggregate collisions, whereas the calculations performed by \citet{DominikTielens1997} are considering particle-particle collisions.
\par
Additionally it should be noted that our $\mathrm{\mu}$m-sized water-ice particles possess a rather broad size distribution, which also influences the sticking threshold, because the smaller particles stick at higher velocities than larger ice grains. Thus, the values derived in this work are to be considered as averages over the size distribution.

\section{Comparison of the experimental results with the collision model by \citet{Krijtetal2013}}\label{Comparion with model}
In this section, we compare our results with the outcome of the collision model developed by \citet{Krijtetal2013}. This collision model describes the energy dissipation in head-on collisions between spherical particles by taking into account adhesion, visco-elasticity and plastic deformation. The only material parameters in this model are the viscous relaxation time $\tau_{\rm vis}$ and the specific surface energy of the material $\gamma$.
\par
We used this model to calculate the coefficient of restitution after the collision of two equal-sized water-ice particles with radii of $r \, = \, 1.5 \, \mathrm{\mu m}$ for different collision velocities. For velocities lower than the sticking threshold, the coefficient of restitution remains zero because sticking occurs, i.e, the velocity after the collision is zero. At higher velocities, the coefficient of restitution reaches values between zero and unity depending on the collision speed, particle size, and material properties. Thus, the highest velocity for which the coefficient of restitution is zero can be regarded as the sticking threshold.
\par
We performed different runs of the collision model to calculate the sticking threshold of the water-ice particles for various values of the viscous relaxation time and the specific surface energy. Both parameters were varied systematically and independently until the model matched the sticking threshold of $9.6 \, \mathrm{m \, s^{-1}}$ that was experimentally derived for temperatures below $\sim 210 \, \mathrm{K}$. We allowed for an uncertainty of $\pm 0.3 \, \mathrm{m\, s^{-1}}$ for the sticking threshold calculated with the collision model. In order to take collisions between ice particles and ice aggregates into account, we changed the mass of the second collision partner to $m_{\rm agg} \, = \, 5 \, m_{\rm 0}$, where $m_{\rm 0}$ is the mass of a water-ice particle with a radius of $1.5\,\mathrm{\mu m}$ \citep[see][and Sect. \ref{Discussion and Outlook}]{Schraepler2011}.
\par
\begin{figure}[t!]
\centering
\includegraphics[angle=0,width=1\columnwidth]{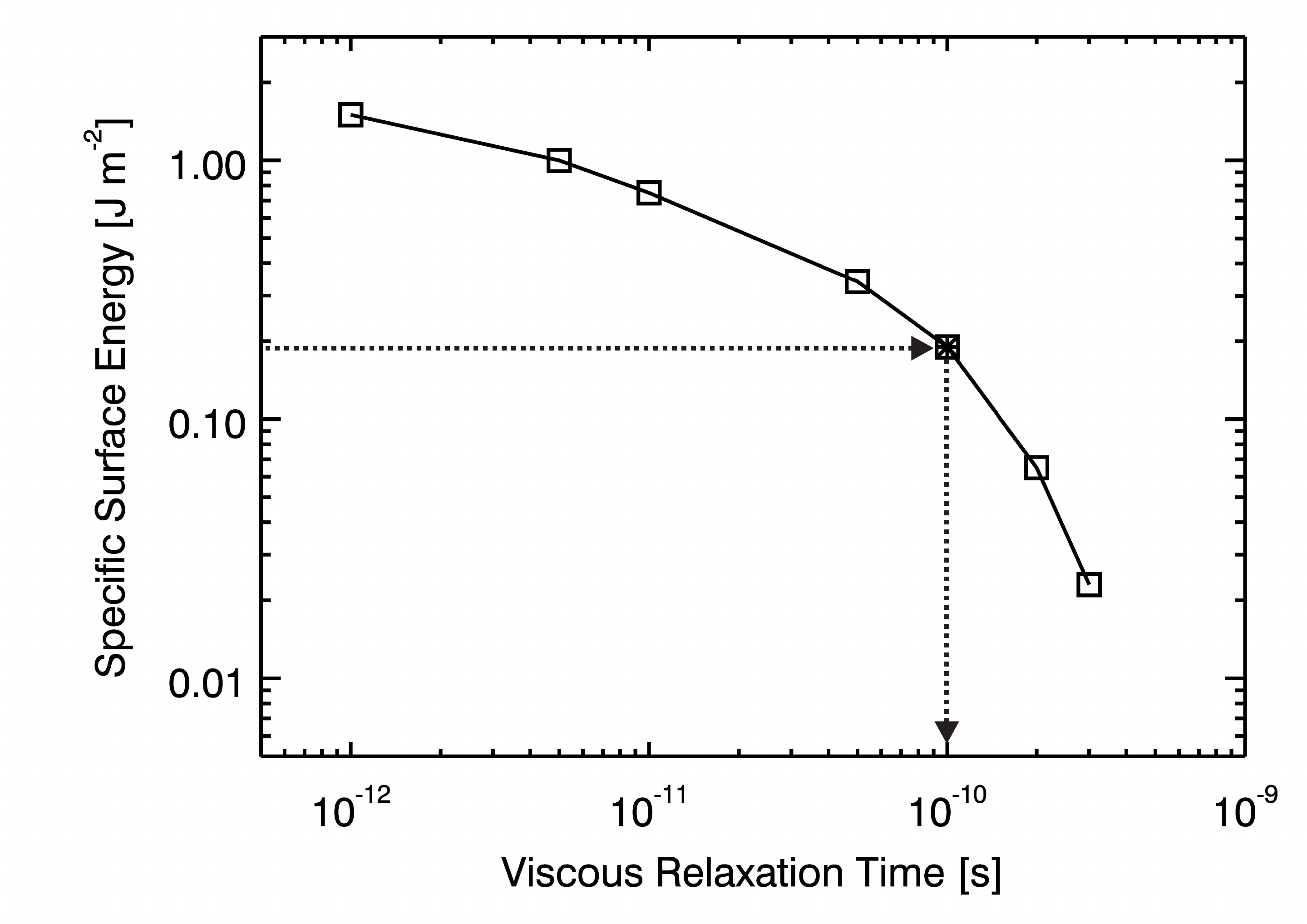}
\put(-167,99.5){$\gamma \, =\, 0.19 \, \mathrm{J \, m^{-2}}$}
\put(-145,32.5){$\tau_{vis} \, =\, 1 \times 10^{-10}\,\mathrm{s}$}
\caption{Combinations of the viscous relaxation time and the specific surface energy (squares) that result in a sticking threshold of $(9.6 \pm 0.3) \, \mathrm{m \, s^{-1}}$ using the collision model by \citet[][squares]{Krijtetal2013}. The asterisk shows the model run performed with a specific surface energy of $\gamma \, = \,  0.19\,  \mathrm{J\, m^{-2}}$ \citep{Gundlach2011b}, which yields a viscous relaxation time of $1 \times 10^{-10}\,\mathrm{s}$ (dotted lines).}
\label{Krijt}
\end{figure}

Fig. \ref{Krijt} shows those combinations of the viscous relaxation time and the specific surface energy that result in a sticking threshold of $(9.6 \pm 0.3) \, \mathrm{m \, s^{-1}}$ (squares). For the model calculations, a Young's modulus of $7 \, \mathrm{GPa}$, a Poisson ratio of $0.25$, and a water-ice mass density of $930 \, \mathrm{kg m^{-3}}$, respectively, were used. One run of the collision model shown in Fig. \ref{Krijt} was performed by assuming a specific surface energy of $\gamma \, = \,  0.19\, \mathrm{J\, m^{-2}}$ (asterisk), a value that was derived from previous experiments performed with the samme $\mathrm{\mu}$m-sized water-ice particles as in this study \citep[see Sect. \ref{Specific surface energy};][]{Gundlach2011b}. In this case, the viscous relaxation time becomes $1 \times 10^{-10}\,\mathrm{s}$ (dotted lines) and matches the range of values $1 \times 10^{-12} \, < \, \tau_{vis} \, < \, 3 \times 10^{-10}$ acquired by fitting other experiments, which were also performed with $\mathrm{\mu}$m-sized particles \citep[see Fig. 8a in][]{Krijtetal2013}.

\section{Discussion and Outlook}\label{Discussion and Outlook}
In the previous Sections, we presented a novel experimental setup with which we grew aggregates consisting of $\mathrm{\mu}$m-sized water-ice particles under astrophysically relevant conditions. We measured the threshold velocities for sticking and erosion in collisions between individual $\mathrm{\mu}$m-sized water-ice particles and aggregates thereof at different temperatures between $114 \, \mathrm{K}$ and $260 \, \mathrm{K}$ (see Sect. \ref{Experimental setup and procedure}). The experiments showed that the sticking threshold is constant at $9.6 \, \mathrm{m \, s^{-1}}$ for temperatures below $\sim 210 \, \mathrm{K}$. At higher temperatures the sticking threshold increases to values between $\sim 30\, \mathrm{m \, s^{-1}}$ and $\sim 50\, \mathrm{m \, s^{-1}}$ (see Fig. \ref{Sticking_Threshold_Temperature_1}). This change in the sticking threshold with increasing temperature is probably caused by the transition from brittle state I to brittle state II at a temperature of $229\, \mathrm{K}$ \citep{Higaetal1998}. In comparison to $\mathrm{\mu}$m-sized silica particles \citep{Poppe2000}, the sticking threshold of water-ice particles is approximately tenfold higher. This dependency was expected due to the increased specific surface energy of water-ice in comparison to silica (see Sect. \ref{Specific surface energy}). The estimated sticking threshold was derived for particle-aggregate collisions, which is the most probable case in protoplanetary disks.
\par
A comparison of our experimental data with the collision model by \citet{Krijtetal2013} provides the opportunity to derive important material properties of $\mathrm{\mu}$m-sized water-ice particles at cryogenic temperatures, i.e, the specific surface energy and the viscous relaxation time (see Sect. \ref{Comparion with model}). Without any restrictions, it is only possible to provide possible combinations of the two material properties (see Fig. \ref{Krijt}). However, as the specific surface energy has been determined in previous laboratory experiments \citep{Gundlach2011b} to be $\gamma \, = \,  0.19\, \mathrm{J\, m^{-2}}$, we derive a viscous relaxation time of $10 \times 10^{-10}\,\mathrm{s}$.
\par
In addition to the sticking threshold, we also estimated the erosion threshold of the $\mathrm{\mu}$m-sized water-ice particles. The experiments showed that the erosion threshold is also temperature independent at low temperatures (below $\sim 210 \, \mathrm{K}$) and reads $15.3 \, \mathrm{m \, s^{-1}}$ (see Fig. \ref{erosion_threshold_temperature}). For comparison, $\mathrm{\mu}$m-sized silica particles possess an erosion threshold of $\sim 2 \, \mathrm{m \, s^{-1}}$ \citep{Schraepler2011,Seizingeretal2013}. This result shows that ice aggregates are more resistant to erosion than silica aggregates. For higher temperatures, the erosion threshold increases to a value of $\sim 50 \, \mathrm{m \, s ^{-1}}$ at a temperature of $(255\pm5)\, \mathrm{K}$.
\par
\begin{figure}[t!]
\centering
\includegraphics[angle=90,width=1\columnwidth]{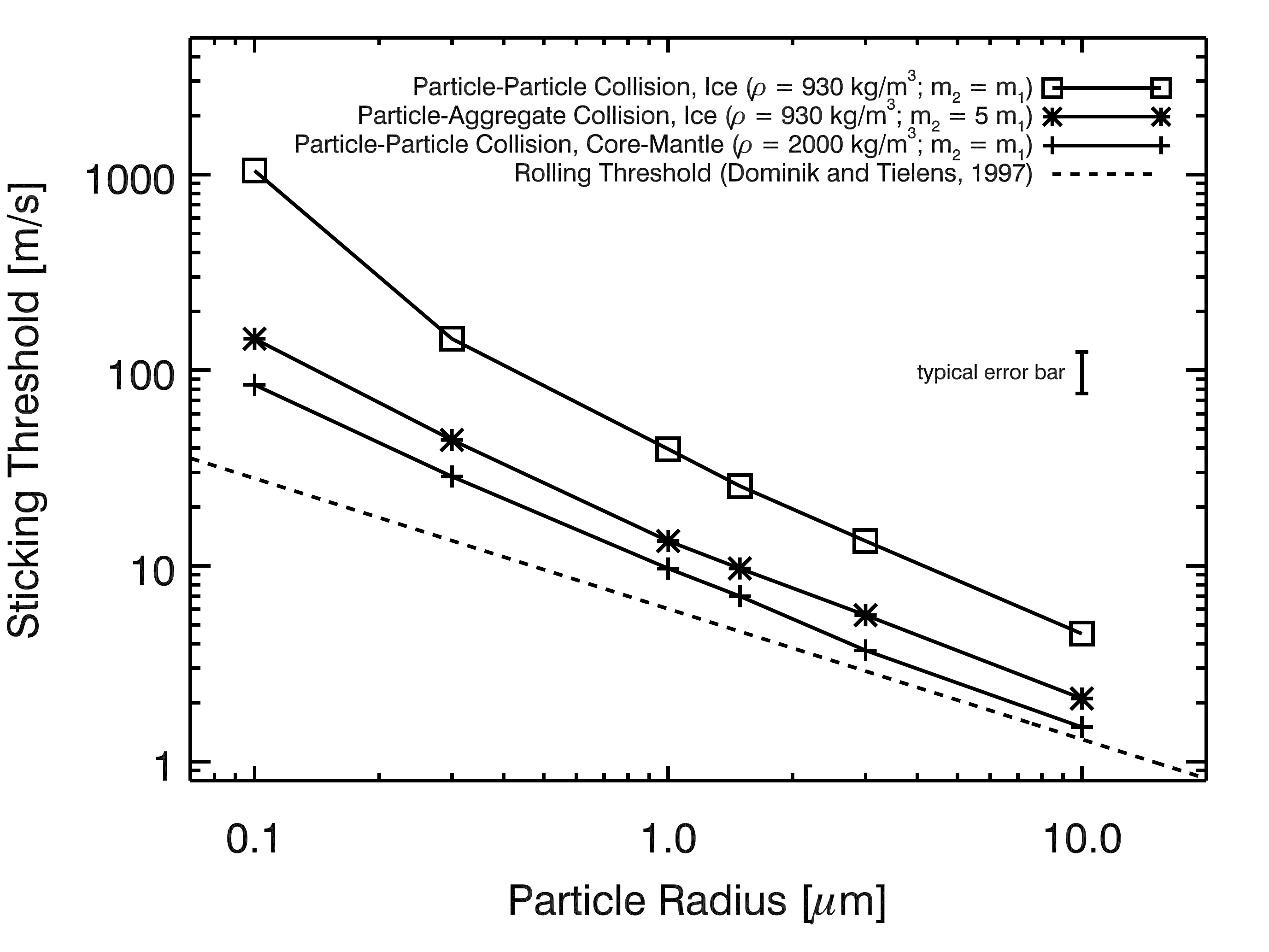}
\caption{Threshold velocity for sticking between two spherical water-ice grains with radii between 0.1\,$\rm \mu m$ and 10\,$\rm \mu m$. The squares show the threshold velocity for sticking if two water-ice monomers collide. Impacts of a single water-ice monomer into a massive aggregate of water-ice particles are shown by the asterisks. If the monomer grains are core-mantle particles with a mass density of $\rho = 2000 \, \mathrm{kg \, m^{-1}}$, the threshold velocity for sticking shifts to the curve with the pluses (in the case of particle-particle collisions). The dashed line indicates the threshold velocity for rolling. The typical error bar shows the model uncertainty arising due to uncertainty of the specific surface energy measurements (see Sect. \ref{Specific surface energy}).}
\label{threshold_velocity_sticking}
\end{figure}

As shown in Sect. \ref{Comparion with model}, our experimental results can be fitted with the collision model by \citet{Krijtetal2013} with reasonable assumptions for the surface energy and the viscous relaxation time (see Fig. \ref{Krijt}). We used this model to predict the threshold velocity for sticking between spherical water-ice particles. For this, we assumed a constant (temperature and particle-size independent) surface energy of $\gamma \, = \,  0.19\, \mathrm{J\, m^{-2}}$ and a viscous relaxation time that is grain-size dependent, as shown in \citet{Krijtetal2013} (their Fig. 8b), and reads
\begin{equation}
\tau_{\rm vis} \, = \, \tau_{\rm vis,\ exp} \, \left(\,\frac{r \, \mathrm{[\mu m]}}{1.5\, \mathrm{\mu m}}\,\right)^{1.11} \, ,
\end{equation}
where $r$ is the particle radius and $\tau_{\rm vis,\ exp} = 10 \times 10^{-10}\,\mathrm{s}$ is the viscous relaxation time derived from the experiments. The results of these calculations are shown in Fig. \ref{threshold_velocity_sticking}, where we plot the threshold velocity for sticking as a function of grain radius for radii between 0.1\,$\rm \mu m$ and 10\,$\rm \mu m$. The three curves show the case of grain-grain collisions (squares), grain-aggregate collisions (asterisks), and grain-grain collisions for core-mantle particles (pluses), respectively. Grain-aggregate collisions were approximated by assigning one of the two colliding grains an effective mass of five times the monomer mass. As \citet{Schraepler2011} showed by comparing their experimental results on grain-aggregate collisions with model predictions by \citet{Konstandopoulos2000}, the target aggregate can be replaced by a single monomer with an effective mass of
\begin{equation}
m_{\rm eff} \, = \, m \, \left(\,1\, +\ C\,n\,\right) \, ,
\end{equation}
with $m$, $C=1$, and $n=4$ being the monomer mass, a rigidity parameter, and the local coordination number. Thus, we get $m_{\rm eff} = 5\, m$. For the core-mantle particles (particle-particle collisions), all characteristic material parameters were kept constant with the exception of the mass density of both grains, which was raised from $\rho = 930 \, \mathrm{kg \, m^{-1}}$ for pure water ice to $\rho = 2000 \, \mathrm{kg \, m^{-1}}$ for core-mantle particles, assuming approximately equal volumes for the water-ice mantle and the silicate core.
\par
Following the collision models by \citet{DominikTielens1997} and \citet{Krijtetal2013}, we can derive the grain size below which collisions are in the hit-and-stick regime, i.e., grains collide without any compaction, and above which growth is expected to occur as compact aggregates. The hit-and-stick threshold is derived by \citet{DominikTielens1997} to be
\begin{equation}
v_{\mathrm{hit-and-stick}} = 5 E_{\mathrm{roll}} \, ,
\end{equation}
where $E_{\mathrm{roll}}$ is the rolling-friction energy. In case of equal-size particles, the rolling-friction energy is given by \citep{DominikTielens1997}
\begin{equation}
E_{\mathrm{roll}} = 3 \pi^2 \gamma r \xi \, ,
\end{equation}
with $\xi$ being the critical displacement before rolling starts. Following \citet{Krijtetal2013}, this critical displacement is given by
\begin{equation}
\xi = \frac{a_{\mathrm{eq}}}{12} \frac{\Delta \gamma}{\gamma} \, ,
\end{equation}
with $a_{\mathrm{eq}}$ being the equilibrium contact radius between the ice grains and $\frac{\Delta \gamma}{\gamma} \approx 1$. The contact radius for equal-size spheres is given by
\begin{equation}
a_{\mathrm{eq}} = \left( \frac{9 \pi \gamma r^2}{4 E^*} \right)^{1/3} \, ,
\end{equation}
with $E^* = 1.5 \times 10^{10} \, \mathrm{Pa}$ being the combined elastic modulus of water ice. Thus, the critical displacement scales as $\xi \propto r^{-2/3}$ and we can calculate below which velocity sticking occurs in the hit-and-stick regime, which yields
\begin{equation}
v_{\mathrm{hit-and-stick}} < \mathrm{min}(v_{\mathrm{stick}},v_{\mathrm{roll}}) \, ,
\end{equation}
with $v_{\mathrm{roll}} = (2 E_{\mathrm{roll}} / m)^{1/2}$ and $m$ being the reduced mass of the projectile and the target particle. This velocity is also shown in Fig. \ref{threshold_velocity_sticking} (dashed line). As can be seen, the threshold velocity for sticking $v_{\mathrm{stick}}$ is always higher than $v_{\mathrm{roll}}$ in the monomer-size range shown in Fig. \ref{threshold_velocity_sticking} and for monomer grains as projectiles. This is not necessarily the case when two ice-aggregates collide, which is, however, not the subject of this study. Thus, water-ice grains grow to highly porous aggregates for velocities below $v_{\mathrm{roll}}$ (dashed line in Fig. \ref{threshold_velocity_sticking}) and to rather compact aggregates above $v_{\mathrm{roll}}$ and below $v_{\mathrm{stick}}$.
\par
The relatively high sticking threshold of water-ice particles found in our experiments is not in contradiction to the idea that icy planetesimals form by direct sticking of sub-$\mathrm{\mu}$m-sized water-ice particles as suggested by \citet{Wada2008,Wada2009,Suyamaetal2008,Suyamaetal2012,Okuzmi2012,Kataokaetal2013}. In this formation scenario, 0.1~$\mathrm{\mu}$m-sized water-ice particles are assumed to stick at velocities up to $60\, \mathrm{m \, s^{-1}}$, which leads to the formation of very porous (with volume filling factors down to $10^{-5}$) ice aggregates. Due to the high porosity, the aggregates are not severely affected by radial drift in the protoplanetary disk and, thus, possess a rather long lifetime before they evaporate. After the formation of the very porous aggregates, gas compression and finally self-gravitational compression lead to a compaction of the planetesimals. So far, this formation scenario was lacking the experimental proof that sub-$\mathrm{\mu}$m-sized water-ice particles do stick at the assumed relatively high velocities. With our measured sticking threshold of $v_{\rm st} \, = \, 9.6 \, \mathrm{m \, s^{-1}}$ for water-ice particles with a mean radius of $r \, = \, 1.5 \, \mathrm{\mu m}$ at low temperatures and applications of the model by \citet{Krijtetal2013}, we can now support this formation scenario under the assumption of very small (sub-$\rm \mu m$) ice particles. As can be seen from Fig. \ref{threshold_velocity_sticking}, a critical sticking velocity of $60\, \mathrm{m \, s^{-1}}$ is reached for water-ice particle radii of $\sim 0.7\, \rm \mu m$ in particle-particle and $\sim 0.2\, \rm \mu m$ for particle-aggregate collisions, respectively. In the case of core-mantle particles, the maximum particle sizes are even smaller. Thus, for maximum particle sizes as indicated above, the assumption of high stickiness of water-ice particles made by \citet{Wada2008,Wada2009,Suyamaetal2008,Suyamaetal2012,Okuzmi2012,Kataokaetal2013} seem to be in agreement with our experimental results and the collision model by \citet{Krijtetal2013}. However, it is at least questionable whether the assumed small particle radii match the ice-particle sizes in protoplanetary disks. Thus, it would be interesting to repeat the model calculations for particles with radii of $\sim 1 \, \mathrm{\mu m}$.
\par
In the future, we plan to perform further laboratory collision experiments with $\mathrm{\mu}$m-sized water-ice particles. Next, we will focus on measuring the coefficient of restitution as a function of ice-particle size and temperature. We intend to derive the individual particle sizes of the impinging water-ice particles from the forward scattered light that enters our long-distance microscope, using Mie scattering theory. These measurements will then allow us to disentangle the specific surface energy and the viscous relaxation time obtained by fitting the experimental data with the collision model developed by \citet{Krijtetal2013}. Performing these future experiments at different temperatures will yield the temperature dependency of the specific surface energy of water ice for temperatures between $\sim 100 \mathrm{K}$ and $\sim 250 \mathrm{K}$. In addition, experiments with core-mantle particles (water-ice particles containing a silicate core) will be performed in order to determine the influence of the silicate cores on the collision properties of $\mathrm{\mu}$m-sized water-ice particles.

\section{Conclusions}\label{Conclusions}
In this paper, we described our experimental work on the deduction of the collision properties of spherical, micrometer-sized water-ice particles under astrophysically relevant conditions, i.e., at low pressures and temperatures. The main results of this work are
\begin{enumerate}
  \item Spherical water-ice particles with a mean radius of $1.5\, \mathrm{\mu m}$ stick in collisions with aggregates of water-ice particles and under astrophysically relevant conditions as long as their impact velocities are below $\sim 10 \, \mathrm{m \, s^{-1}}$ (see Sect. \ref{The sticking and the erosion threshold of micrometer-sized water-ice particles} and Figs. \ref{growth_1}, \ref{growth_2}, \ref{sticking_comp} and \ref{Sticking_Threshold_Temperature_1}). The threshold velocity for sticking is about tenfold higher than that of spherical $\rm SiO_2$ particles with comparable size \citep{Poppe2000}, which can be explained by a tenfold higher specific surface energy of water ice.
  \item The threshold velocity for sticking of  $1.5\, \mathrm{\mu m}$ water-ice particles is temperature independent for $T < 210\,\mathrm{K}$ and increases with increasing temperature for $T > 210 \, \mathrm{K}$ (see Fig. \ref{Sticking_Threshold_Temperature_1}). As all astrophysically relevant water-ice temperatures in molecular clouds and protoplanetary disks are well below $210 \,\mathrm{K}$ \citep{BerginTafalla2007,LecarEtal2006}, laboratory experiments on the collision and sticking behavior of water-ice particles should be done at temperatures below $210 \,\mathrm{K}$ to avoid increased stickiness, or sintering effects.
  \item The threshold velocity for sticking of the  $1.5\, \mathrm{\mu m}$ water-ice particles can be reconstructed with the collision model of \citet{Krijtetal2013} with reasonable assumptions for the specific surface energy of water ice of $\gamma \, = \,  0.19\, \mathrm{J\, m^{-2}}$ \citep{Gundlach2011b} and the viscous relaxation time of $\tau_{\mathrm{vis}} \, =\, 1 \times 10^{-10} \, \mathrm{s}$, respectively. Applying the model by \citet{Krijtetal2013} allows the deduction of a threshold velocity for sticking in monomer-monomer collisions of $\sim 25 \, \mathrm{m \, s^{-1}}$ (see Fig. \ref{threshold_velocity_sticking}). Varying the size of water-ice monomers in monomer-aggregate collisions gives sticking thresholds between $\sim 150 \, \mathrm{m \, s^{-1}}$ for grain radii of $0.1\, \mathrm{\mu m}$ and $\sim 2 \, \mathrm{m \, s^{-1}}$ for grain radii of $10\, \mathrm{\mu m}$ (see Fig. \ref{threshold_velocity_sticking}). Core-mantle particles possess threshold velocities of 60\% to 70\% of the values of pure ice particles (particle-particle collisions; see Fig. \ref{threshold_velocity_sticking}).
  \item For impact velocities exceeding the sticking threshold, the water-ice aggregates become eroded (see Fig. \ref{erosion_1}). The onset of erosion occurs typically for velocities above 1.5 times the upper limit for sticking (see Figs. \ref{Sticking_Threshold_Temperature_1} and \ref{erosion_threshold_temperature}). This finding is in agreement with predictions by the collision model of \citet{DominikTielens1997}.
\end{enumerate}

\acknowledgments
This project was supported through the DFG priority programm "Physics of the Interstellar Medium" under grant BL 298/19-1. We thank Sebastiaan Krijt, Carsten Güttler, and Daniel Heißelmann for programming their collision model in IDL and Florian Heimbach for providing the sketch of the experimental setup. Furthermore, we thank the anonymous reviewer for detailed and helpful comments, which helped improving this manuscript considerably.

\bibliography{bib}
\bibliographystyle{apj}

\end{document}